\documentclass[fleqn,usenatbib]{mnras}

\usepackage{newtxtext,newtxmath}
\usepackage[T1]{fontenc}

\DeclareRobustCommand{\VAN}[3]{#2}
\let\VANthebibliography\thebibliography
\def\thebibliography{\DeclareRobustCommand{\VAN}[3]{##3}\VANthebibliography}

\usepackage{graphicx}	% Including figure files
\usepackage{amsmath}	% Advanced maths commands
\graphicspath{{figures/}}
\usepackage{url}
\usepackage{multirow}
\newcommand{\Msun}{\;M$_\odot$}
\newcommand{\molh}{H$_{2}$}	% per cm-squared

\newcommand{\gureft}{\textsc{gureft}}

\defcitealias{Finkelstein2022b}{FB23}
\defcitealias{Finkelstein2023}{FL23}

\title[Ultra-high-z galaxies in the SC SAM]{Are the ultra-high-redshift galaxies at $z >10$ surprising in the context of standard galaxy formation models?}

\author[L. Y. A. Yung et al.]{L. Y. Aaron\ Yung,$^{1}$\thanks{E-mail: aaron.yung@nasa.gov}\thanks{NASA Postdoctoral Fellow}
Rachel S.\ Somerville,$^{2}$
Steven L.\ Finkelstein,$^{3}$
Stephen M.\ Wilkins,$^{4,5}$\newauthor
Jonathan P.\ Gardner$^{1}$
\\
$^{1}$Astrophysics Science Division, NASA Goddard Space Flight Center, 8800 Greenbelt Rd, Greenbelt, MD 20771, USA\\
$^{2}$Center for Computational Astrophysics, Flatiron Institute, 162 5th Ave, New York, NY 10010, USA\\
$^{3}$Department of Astronomy, The University of Texas at Austin, Austin, TX, USA\\
$^{4}$Astronomy Centre, University of Sussex, Falmer, Brighton BN1 9QH, UK\\
$^{5}$Institute of Space Sciences and Astronomy, University of Malta, Msida MSD 2080, Malta
}

\date{Accepted XXX. Received YYY; in original form ZZZ}

\pubyear{\the\year{}}

\begin{document}
\label{firstpage}
\pagerange{\pageref{firstpage}--\pageref{lastpage}}
\maketitle

\begin{abstract}
A substantial number of ultra-high redshift ($8 \lesssim z \lesssim 17$) galaxy candidates have been detected with \textit{JWST}, posing the question: are these observational results surprising in the context of current galaxy formation models? We address this question using the well-established Santa Cruz semi-analytic models, implemented within merger trees from the new suite of cosmological $N$-body simulations \gureft, which were carefully designed for ultra-high redshift studies. Using our fiducial models calibrated at $z=0$, we present predictions for stellar mass functions, rest-frame UV luminosity functions, and various scaling relations. We find that our (dust-free) models predict galaxy number densities at $z\sim 11$ ($z\sim 13$) that are an order of magnitude (a factor of $\sim 30$) lower than the observational estimates. We estimate the uncertainty in the observed number densities due to cosmic variance, and find that it leads to a fractional error of $\sim$20-30\% at $z=11$ ($\sim$30--80\% at $z=14$) for a 100 arcmin$^{2}$ field. We explore which processes in our models are most likely to be rate-limiting for the formation of luminous galaxies at these early epochs, considering the halo formation rate, gas cooling, star formation, and stellar feedback, and conclude that it is mainly efficient stellar-driven winds. We find that a modest boost of a factor of $\sim 4$ to the UV luminosities, which could arise from a top-heavy stellar initial mass function, would bring our current models into agreement with the observations. Adding a stochastic component to the UV luminosity can also reconcile our results with the observations.
\end{abstract}


%%%%%%%%%%%%%%%%%%%%%%%%%%%%%%%%%%%%%%%%%%%%%%%%%%
%%%%%%%%%%%%%%%%% BODY OF PAPER %%%%%%%%%%%%%%%%%%

\section{Introduction}

Understanding the physical processes that shape galaxies is one of the most important and challenging open problems in astrophysics. Over the past decades, advances in instrumentation have enabled us to push the ``redshift frontier'' back ever further in distance and cosmic time, discovering objects that formed in the very early Universe, shortly after the Big Bang. Before the launch of \textit{JWST}, the highest redshift reported galaxy candidates were at redshifts of $\sim 9$--11, corresponding to formation times about 400-500 Myr after the Big Bang \citep[e.g.][]{Oesch2016, Finkelstein2021}. 

After many years of anticipation, the \textit{JWST} \citep{Gardner2006, Gardner2023} was launched in December, 2021. With its 6.5m aperture and infrared sensitivity, \textit{JWST} was designed to be capable of detecting extremely-distant, high-redshift galaxies. Almost as soon as the first NIRCam images from Early Release Observations (ERO) and Early Release Science (ERS) such as SMACS0723, GLASS, and CEERS were released, reports of galaxy candidates that broke the existing redshift record started to appear \citep[e.g.][]{Finkelstein2022a,Naidu2022,Castellano2022,Atek2022,Adams2022,Donnan2022,Harikane2022}. Additionally, \citet{Labbe2022} reported the discovery of several galaxies at $z\sim 8$--10 with extremely large stellar mass estimates, suggesting that the formation of these galaxies' progenitors must have begun very early. 

Our current state-of-the-art models of galaxy formation are set within the paradigm of the $\Lambda$ Cold Dark Matter ($\Lambda$CDM) model for structure formation \citep{Blumenthal1984}. In this picture, galaxies form within dark matter halos, and gas can only cool and form stars within halos above a critical mass corresponding to the temperature where cooling becomes possible. Thus, the epoch of formation of the first stars and galaxies is largely determined by the evolution of the dark matter halo mass function, which is set by the primordial power spectrum of density fluctuations and gravitational collapse.  The very first stars to form out of the dark ages are generally referred to as Pop III stars, and they are formed from metal-free gas which must rely on \molh\ as a coolant. These Pop III stars are expected to start forming at around $z\sim 30$--20 in 3-4 $\sigma$ peaks of the density field, corresponding to ``mini''-halos with masses of $10^{5}-10^{6}$ \Msun\ \citep{Haiman1997,Barkana2001, Bromm2002, Visbal2018}. 
Pop III stars quickly enrich the gas in their surroundings to the point where the dominant mode of star formation becomes Pop II stars, which form in more massive ``atomic cooling'' halos, with masses of $10^{7}-10^{8}$ \Msun\ \citep{Oh2002}. 

The discovery of these ultra-high redshift galaxy candidates initially appeared surprising for several reasons. These early data are relatively shallow, implying that these candidates are fairly luminous. Yet, at the same time, the areas covered are quite small ($\sim$ a few tens of arcmin$^2$). Additionally, the newly detected objects imply very weak evolution in the bright end of the luminosity function from $z\sim 12$ to $z\sim 4$ \citep{Donnan2022, Finkelstein2022b, Harikane2022,Finkelstein2023}, while the halo mass function evolves very strongly at the massive end over this interval \citep[e.g.][]{Behroozi2020}. Some published semi-analytic models \citep{Dayal2014, Dayal2019,Yung2020a} and hydrodynamic simulations \citep{Wilkins2022} predict many fewer luminous galaxies at $z\gtrsim 10$ than the new observations suggest \citep[see e.g.][their Fig. 15]{Harikane2022}. Some empirical models extrapolated from lower redshifts suffer similar issues \citep[e.g.][]{Behroozi2020,Mirocha2022}.

One of the first questions that arose following these discoveries was whether the existence of these early galaxies violates basic constraints placed by the abundance of dark matter halos in standard $\Lambda$CDM. \citet{Labbe2022} conducted a search for high redshift galaxies with a strong Balmer break, and discovered objects at $z\gtrsim 6$ with stellar masses that seemed surprisingly high ($10^{10}-10^{11} M_{\odot}$). Based on the \citet{Labbe2022} results that were originally posted on the archive (arXiv:2207.12446v1), \citet{Boylan-Kolchin2022} argued that these galaxies were several orders of magnitude more numerous than expected in $\Lambda$CDM, and violated constraints on the total amount of baryonic mass available in standard cosmologies. \citet{Lovell2022} performed an Extreme Value Statistics analysis on several of the available samples, again adopting the estimated stellar masses of the original \citet{Labbe2022} analysis, and found that these objects were in moderate ($3 \sigma$) tension with standard $\Lambda$CDM. It is important to note that these studies did not account for uncertainties from field-to-field variance due to large scale structure. Additionally, following an in-flight update to the calibration of the NIRCam instrument released on July 29, 2022, the stellar mass estimates in the \citet{Labbe2022} sample were revised, leading to a removal of any fundamental tension with $\Lambda$CDM, as reflected in the revised version of \citet{Boylan-Kolchin2022}. 

\citet{Mason2022} showed that a ``maximally optimistic'' empirical model, in which star formation is 100\% efficient, all gas in halos is made available to form stars, and all galaxies have young ages ($10$ Myr), is able to produce rest-UV luminosity functions four orders of magnitude higher than current observations, again implying that there is no fundamental tension with $\Lambda$CDM. Dust could play a role in causing the unexpectedly weak redshift evolution in the UV LF from $z\sim 15$--8 \citep{Ferrara2022}. 

In interpreting these results, it is important to keep in mind that there are still significant observational uncertainties on the true number densities of ultra-high redshift galaxies. These galaxy candidates have been identified through the Lyman-break selection techniques or via photometric redshifts. Some of the objects that are currently being counted as $z>10$ galaxies may actually be at much lower redshifts. Observations with the NIRSpec instrument on \textit{JWST} have so far suggested that a fairly high fraction of photometric candidates at $z\sim 7$--13 are indeed at redshifts close to those implied by the photometric redshift analysis \citep{Curtis-Lake2022, ArrabalHaro2023a, Fujimoto2023}. 
However, in some cases the photo-z analysis can yield a strongly bimodal probability distribution, such as the objects with one peak at at $z\sim 5$ and another at $z\sim 16$--18 discussed by \citet{Naidu2022} and \citet{Zavala2022}. Recently, NIRSpec spectroscopy revealed that a bright ($m_{\rm F277W}=26.5$) object in the CEERS field that was identified as a fairly robust candidate for being at $z\sim 16.4$ \citep{Donnan2022,Naidu2022a,Harikane2022,Finkelstein2022b,Bouwens2022b} instead has a redshift of $4.9$ \citep{ArrabalHaro2023a}. 
Furthermore, our understanding of the calibration of \textit{JWST}'s instruments is still an evolving process, and field-to-field variance is expected to be significant for the very small areas that have been probed up until now.  A further point is that stellar mass estimates based on a few photometric bands have very large uncertainties \citep{Cullen2022}.

As many of the studies mentioned above rely on estimates of the dark matter halo mass function and halo growth rates, it is worth asking how accurately these are known over the mass and redshift ranges of interest here. The answer is that up to now, the constraints on this fundamental quantity for an up to date cosmology are surprisingly poor. \citet{Mason2022} use mass functions from  \citet{Reed2007}, which adopts values of the cosmological parameters that are not consistent with results from the Planck satellite  \citep{Planck2016, Planck2018}. Other studies (such as \citet{Boylan-Kolchin2022} and \citet{Ferrara2022}) adopt halo mass functions from the \citet{Sheth1999} or \citet{Sheth2002} models, which can disagree with $N$-body based results by up to an order of magnitude at high redshift. Others extrapolate fits to $N$-body simulations that were calibrated at much lower redshifts \citep{Lovell2022,Mirocha2022}.
The $z > 10$ UV LF predictions presented in \citet{Yung2020a} adopted $z = 11$ to 15 HMFs that were fitted simultaneously to both a large cosmological simulation \citep[Bolshoi-Planck with $250$ Mpc h$^{-1}$ on a side;][]{Klypin2016} and a high-resolution, smaller volume cosmological simulation \citep{Visbal2018}, and relied on merger trees based on the Extended Press-Schechter formalism, which are not well tested at these early epochs.
\emph{None of these recent studies make use of dark matter halo mass functions and merger trees robustly measured from $N$-body simulations with up to date cosmological parameters over the relevant halo mass and redshift ranges}. 

Semi-analytic models (SAMs) have proven an invaluable tool for \textit{forecasting} observations for high to ultra-high redshift preceding the launch of \textit{JWST} and  for interpreting JWST results \citep{Qin2017, Qin2023, Dayal2019, Yung2019, Hutter2020}. The computational efficiency of this modelling approach, in contrast to conventional numerical methods, makes it possible to simulate galaxies in the ultra-high-redshift universe that are both low mass and extremely rare, which is difficult to achieve with a single numerical hydrodynamic simulation. Furthermore, it enables the interpretation of \emph{JWST} observations in terms of \emph{physical processes}, which is not the case with empirical models (e.g. halo occupation distribution or sub-halo abundance matching).

In this work, we exploit a new suite of dissipationless $N$-body simulations that were designed specifically as a backbone for semi-analytic models of galaxy formation to make predictions for the ultra-high redshift Universe. Gadget at Ultrahigh Redshift with Extra-fine Timesteps (\gureft, pronounced \emph{graft}), is a suite of four cosmological boxes, spanning a range of particle mass from $1.5 \times 10^4$ to $8.5 \times 10^7$ \Msun\ and volumes 5 to 90 Mpc h$^{-1}$ on a side. The adopted cosmological parameters are consistent with constraints from the Planck satellite \citep{Planck2016} and the ones adopted by the MultiDark simulation suite \citep{Klypin2016}. Moreover, most existing $N$-body simulations with publicly available halo catalogues and merger trees have only a small number of snapshots saved at very early times corresponding to $z\gtrsim 6$. In \gureft, many snapshots are saved between $z\sim 40$ to 6, such that we ensure that snapshots are saved every tenth of the typical halo dynamical time. By \emph{grafting} together the results of these different boxes, we can obtain uniquely robust constraints on the assembly histories and abundances of dark matter halos at these early times. 

We couple these new merger trees with a well established model for galaxy formation, the Santa Cruz semi-analytic model \citep{Somerville1999, Somerville2008, Somerville2015}. It contains standard prescriptions for gas cooling, star formation, chemical enrichment, stellar feedback, and other processes. The models are calibrated to match $z=0$ observations, and are not re-calibrated at other redshifts. The fiducial models have previously been shown to match observations of galaxy rest-UV luminosity functions and other observables out to $z\sim 10$ \citep{Somerville2015,Yung2019,Yung2019a,Yung2021}. 
Here we present updated, much more robust predictions for the properties of the galaxy population at the new high redshift frontier $z\sim 8$--17. We forward model our predicted star formation and chemical enrichment histories into the observational space of rest-UV luminosity and compare with the observations in that plane, in order to avoid having to make uncertain assumptions about stellar masses. 

The structure of the paper is as follows. In Section~\ref{sec:scsam}, we present a brief overview of the physical ingredients in the Santa Cruz semi-analytic models. In Section~\ref{sec:gureft}, we present a brief overview of the \gureft\ simulation suite. In Section~\ref{sec:results}, we present our main results. We discuss the implications of our results in Section~\ref{sec:discussion} and summarize and conclude in Section~\ref{sec:conclusions}.

\section{The physical modelling pipeline}
In this section, we provide a concise summary of the Santa Cruz semi-analytic model for galaxy formation (Section \ref{sec:scsam}) and briefly describe the new suite of $N$-body simulations and dark matter halo merger trees used in this work (Section \ref{sec:gureft}). Both the Santa Cruz SAM and \gureft\ simulation suite adopt cosmological parameters $\Omega_m = 0.308$, $\Omega_\Lambda = 0.693$, $H_0 = 67.8$ km s$ ^{-1}$ Mpc$ ^{-1}$, $\sigma_8 = 0.823$, and $n_s = 0.96$; which are broadly consistent with the ones reported by the Planck Collaboration in 2015 (Planck Collaboration XIII \citeyear{Planck2016}), and the same as those adopted by the Bolshoi-Planck simulation from the MultiDark suite \citep{Klypin2016}.

\subsection{Santa Cruz semi-analytic model for galaxy formation}
\label{sec:scsam}
The semi-analytic model (SAM) developed by the Santa Cruz group is a versatile galaxy formation model that implements a suite of carefully curated physical processes to track the formation and evolution of galaxies via dark matter halo merger trees \citep{Somerville1999, Somerville2008, Somerville2012, Somerville2015, Somerville2021, Popping2014}. We refer the reader to these papers for a full description of the model components and \citet{Yung2022} for a schematic flowchart of the internal workflow of the model. The well-established model has been shown to reproduce the observed evolution of distribution functions of galaxy properties at $0 < z < 6$ \citep{Somerville2015} and $4 < z < 10$ \citep{Yung2019, Yung2019a}, with free parameters in the model only calibrated to a subset of observational constraints at $z\sim0$.

The Santa Cruz SAM includes a suite of \textit{standard} physical processes that are also adopted by other semi-analytic models or incorporated in cosmological hydrodynamic simulations. These processes include cosmological accretion and atomic cooling, suppression of gas accretion after reionization by the meta-galactic photoionizing background, star formation and stellar-driven winds, chemical evolution, black hole feedback, and mergers. The cold gas disc is partitioned into atomic, molecular, and ionized components depending on the gas-phase metallicity of the disc using fitting functions based on numerical hydrodynamic simulations by \citet[][denoted as GK]{Gnedin2011}. 
In addition, the model adopts a \molh-based star formation recipe. In a similar spirit as the Kennecutt-Schmidt star formation (SF) relation where SFR is proportional to the surface density of cold gas \citep[e.g.][]{Kennicutt1998}, we adopt a SF relation that scales with the surface density of molecular hydrogen (e.g. $\Sigma_\text{SFR} \propto \Sigma_\text{\molh}^\alpha$), where the slope $\alpha$ steepens from 1 to 2 above a critical value of $\Sigma_\text{\molh}$. This steepening is motivated by observations \citep{Sharon2013,Rawle2014, Hodge2015, Tacconi2018} as well as theory \citep{Ostriker2010}, and has been shown to be essential in order for the SAM 
to reproduce the observed evolution in the $z>4$ UV luminosity functions, stellar mass functions and star formation rate distribution functions \citep{Somerville2015,Yung2019}. 

The model parameters are calibrated as shown in \citet{Gabrielpillai2022}, such that the SAM outputs match the observed $z\sim0$ stellar mass function \citep{Baldry2012, Bernardi2013, Moustakas2013}, stellar-to-halo mass ratio \citep{Rodriguez-Puebla2017},  cold ISM gas fraction vs. stellar mass \citep{Calette2018, Catinella2018}, stellar metallicity \citep{Gallazzi2005, Kirby2011a}, and $M_\text{BH}$--$M_\text{bulge}$ relation \citep{McConnell2013, Kormendy2013}. These physical parameters are not re-\textit{`tuned'} at higher redshift, and we use identical values of the parameters in this study. Note that these parameters are slightly different from the ones used in the Yung et al. \textit{JWST} forecast paper series, due to the switch from Extended Press-Schecter based merger trees to N-body based trees. 

The full star formation and chemical enrichment history is convolved with publicly available results from the stellar population synthesis (SPS) model Binary Population and Spectral Synthesis \citep[BPASS\footnote{\url{https://bpass.auckland.ac.nz/}, v2.3};][]{Eldridge2017, Byrne2022} to produce a high-resolution SED for each galaxy. 
The rest-frame UV magnitude is calculated from the SED with a tophat filter centred around 1600\AA.
\citet{Yung2020} and \citet{Yung2020a} explored various SPS models and have shown that the binary star populations in BPASS yield a higher ionizing photon budget which may help reproduce the reionization history implied by various constraints derived from intergalactic medium and cosmic microwave background observations (see \citealt{Robertson2013} and \citealt{Robertson2015} for a thorough review on the various types of constraints and how they are derived). However, at rest 1600\AA\ the BPASS models yield very similar predictions to single star models such as those of \citet{Bruzual2003}, and we do not expect our predictions to be very sensitive to the assumed stellar population models.
The version of the BPASS model adopted in this work assumes a fiducial broken power law stellar Initial Mass Function, with an upper slope $a_1 = -1.30$ between $0.1$ -- $0.5$ \Msun\ and a lower slope $a_2 = -2.35$ between $0.5$ -- $300$ \Msun. In this work, we utilize SEDs with stellar metallicity spanning $Z = 10^{-5}$ to 0.040.
We note that dust attenuation is omitted in this work, as we wish to obtain the most optimistic predictions for the ultra-high-$z$ galaxy populations. In addition, we include only stellar continuum emission, and neglect nebular emission, although we will include the latter in a forthcoming work.

\begin{figure}
    \includegraphics[width=1\columnwidth]{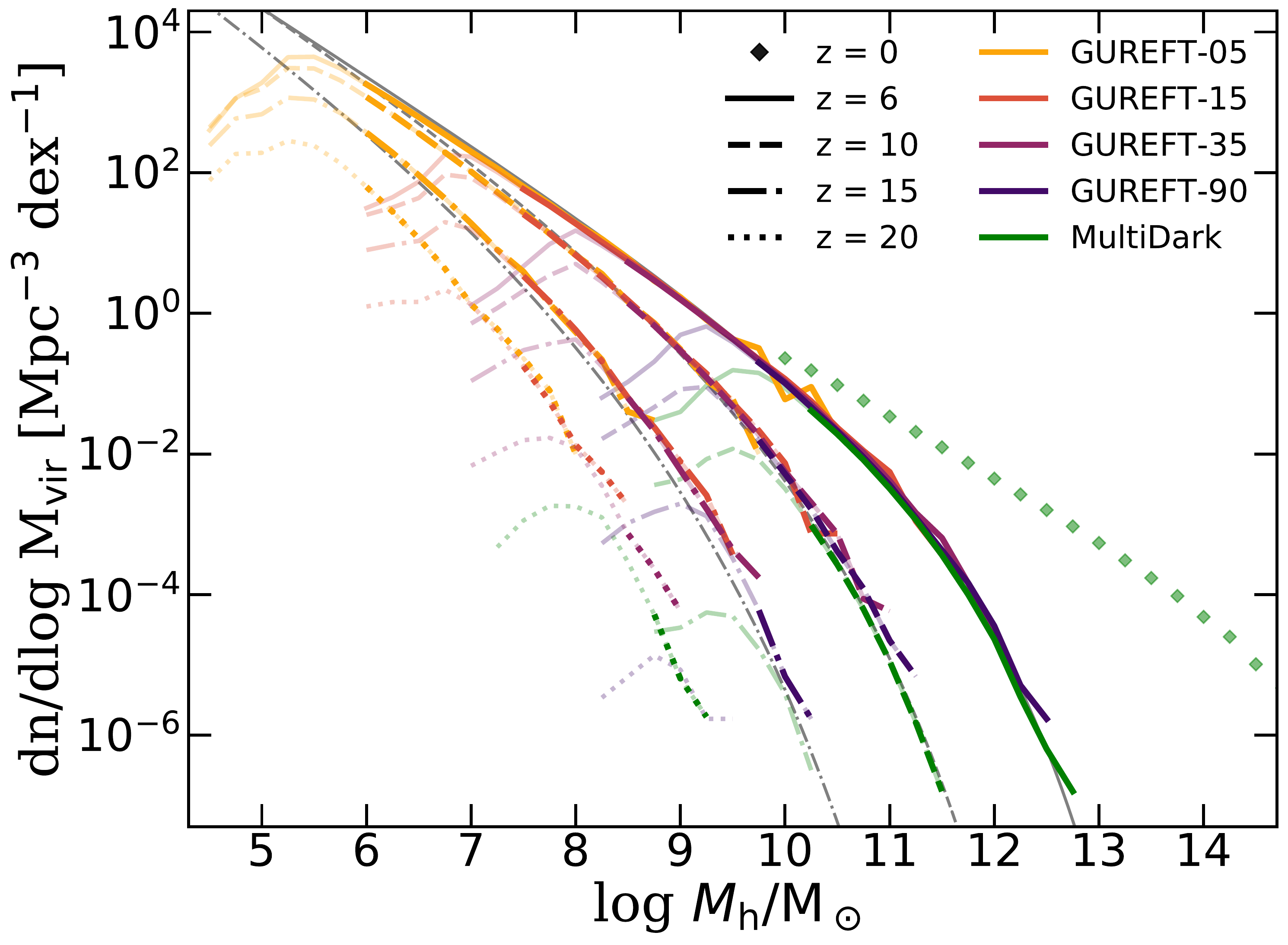}
    \caption{
        Halo mass functions from the suite of \gureft\ boxes at $z = 6$, 10, 15, and 20 \citep{Yung2023b}. Boxes with different volumes are represented by different colours and redshift is represented by different line styles. For comparison, we also show HMFs from the MultiDark simulation suite for comparison at $z = 6$, 10, 15 (from Bolshoi-Planck), and 20 (from VSMDPL), shown with green lines with matching line styles for redshift \citep{Klypin2016}. 
        The line colour fades for individual HMF for the mass range that falls below 50 DM particles for the respective simulation to indicate where the HMF begins to become incomplete.
        We show the $z=0$ HMF from the Bolshoi-Planck simulation with green diamonds.
        The grey lines show the HMFs adopted in previous work \citep[up to $z=15$, e.g.][]{Yung2019, Yung2020a}, fitted to \citet{Rodriguez-Puebla2016} and \citet{Visbal2018}.
    }
    \label{fig:gureft_hmf}
\end{figure}

\subsection{GUREFT: A new suite of N-body simulations and merger trees for the ultra-high redshift Universe}
\label{sec:gureft}
Gadget at Ultrahigh Redshift with Extra-Fine Timesteps (\gureft, pronounced \textit{graft}) is a suite of dark matter-only cosmological simulations designed to have mass and temporal resolution sufficient to capture the merger history of dark matter halos at extreme redshifts \citep{Yung2023b}. 
The merger trees extracted from this simulation suite will ultimately be \textit{grafted} together to paint a holistic picture of the earliest emergence of halos that are massive enough to host galaxies, which are not available from most publicly accessible cosmological simulations due to insufficient mass resolution and sparsely spaced snapshots (see Table \ref{tab:snaptable}).
The suite consists of four simulated volumes, with 5, 15, 35, and 90 Mpc\,$h^{-1}$ on a side, containing $1024^3$ dark matter particles, with mass resolutions specified in Table~\ref{tab:snaptable}.
These boxes are labelled \gureft-05, \gureft-15, \gureft-35, \gureft-90 accordingly.
The initial conditions for the suite are generated with the MUlti-Scale Initial Conditions \citep[\textsc{Music,}][]{Hahn2011} code, with second-order Lagrangian perturbation theory (2LPT) at $z = 200$. 
The evolution of dark matter is carried out with the publicly available version of \textsc{Gadget}-2 code\footnote{\url{https://wwwmpa.mpa-garching.mpg.de/gadget/}} \citep{Springel2005a} down to $z = 6$.

\begin{table}
	\centering
	\caption{A summary of ultra-high-redshift resources available in current state-of-the-art cosmological simulations, including the highest redshift where snapshots are stored, the number of snapshots available at $z \gtrsim 6$, and the dark matter particle mass.
    }
	\label{tab:snaptable}
	\begin{tabular}{lcccc}
    \hline
    Simulation &  highest $z$    & snapshots  & $M_\text{DM particle}$  & Volume   \\
               &  snapshot & at $z\gtrsim6$  & [M$_\odot$]  & [Mpc$^3$]         \\
    \hline
    VSMDPL           & 24.98 & 62  &  $9.1\times 10^6$    & $236.0^3$ \\
    Bolshoi-Planck   & 16.98 & 29  &  $2.3\times 10^8$    & $368.7^3$ \\
    TNG100-1-Dark    & 20.05 & 14  &  $8.9\times 10^6$    & $75.0^3$  \\
    \hline
    \hline
    \multicolumn{3}{l}{the \gureft\ simulation suite}\\
    \hline
    \gureft-05 & \multirow{4}{4em}{\,\,\,\,\;40.00} & \multirow{4}{4em}{\,\,\,\,\,\,\;171} & $1.5 \times 10^4$ &   $7.4^3$\\
    \gureft-15                                                                           &&& $4.0 \times 10^5$ &  $22.1^3$\\
    \gureft-35                                                                           &&& $5.0 \times 10^6$ &  $53.6^3$\\
    \gureft-90                                                                           &&& $8.5 \times 10^7$ & $132.0^3$\\
 \hline
	\end{tabular}
 \end{table}

For each simulated volume, 170 snapshots are saved between $z = 40$ to 6 with spacing roughly one-tenth of the typical halo dynamical time.
For comparison, the Bolshoi-Planck simulation only has 30 snapshots and the IllustrisTNG simulations \citep{Nelson2019} has 14 snapshots in the same redshift range.
Halo and merger tree catalogues are generated with the \textsc{rockstar} \citep{Behroozi2013b} and \textsc{consistent-tree} codes \citep{Behroozi2013c}, based on the virial mass definition of \citet{Bryan1998}.

Fig.~\ref{fig:gureft_hmf} shows the halo mass functions (HMFs) between $z \sim 6$ to 20 from the four \gureft\ boxes (colour-coded). We also show predictions from the Bolshoi-Planck simulations at $z = 6$, 10, and 15 for comparison, as well as the $z=0$ halo mass function to guide the eye. 
We note that no mass cutoff is applied in the making of these HMFs. Thus, the \textit{turnover} observed on the low-mass end reflects where the halo populations become incomplete due to the mass resolution of the simulation, which is typically $\sim 100$ times larger than the mass of DM particles.
Tabulated HMFs constructed by combining well-resolved halos from the four \gureft\ volumes are available in Appendix \ref{sec:tabulated}. For more details on the \gureft\ simulations, as well as a comparison of the halo mass functions and other properties with past work and commonly used analytic models and fitting functions, please see \citet{Yung2023b}.

\section{Results}
\label{sec:results}

In this section, we show predicted properties of the galaxy population simulated with the Santa Cruz SAM with merger trees from the suite of \gureft\ simulations.

\begin{figure*}
    \includegraphics[width=2\columnwidth]{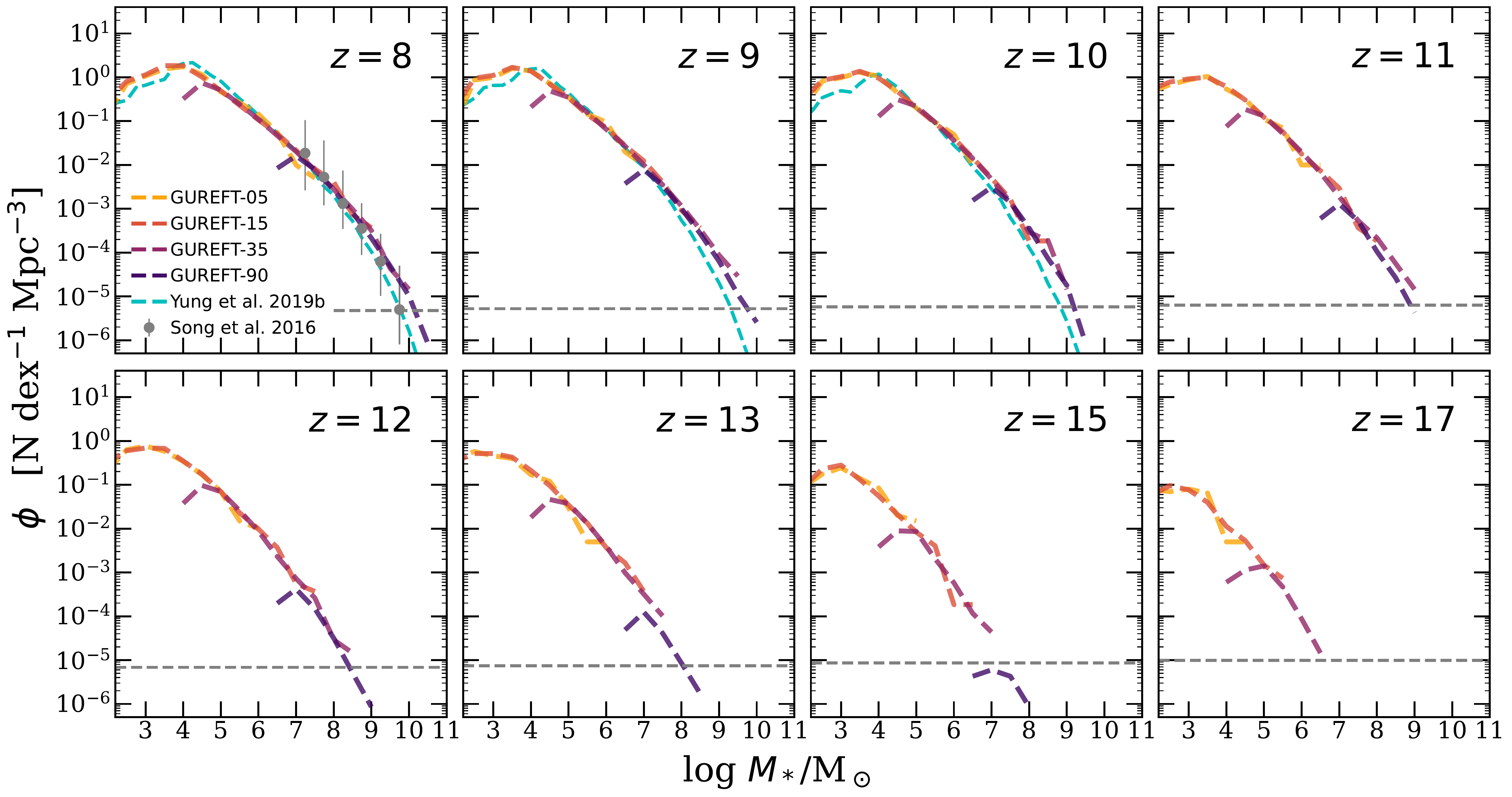}
    \caption{
        Predicted high- to ultrahigh-redshift stellar mass functions from the Santa Cruz SAM using \gureft\ merger trees. The stellar mass functions based on merger trees extracted from the four different resolution \gureft\ boxes are shown with different colours, as indicated on the figure label. Results are compared to the estimated stellar mass function based on \textit{HST} observations reported by \citet{Song2016} at $z \lesssim 8$, and previously published models from \citet[][cyan dashed lines]{Yung2020a}.
        The grey dashed horizontal line marks where one object is expected for a survey area similar to that of the full CEERS area ($\sim 100$ arcmin$^2$) with $\Delta z = 1$ centred around the redshift labelled for each panel.
        }
    \label{fig:gureft_SMF_allz}
\end{figure*}

\begin{figure*}
    \includegraphics[width=2\columnwidth]{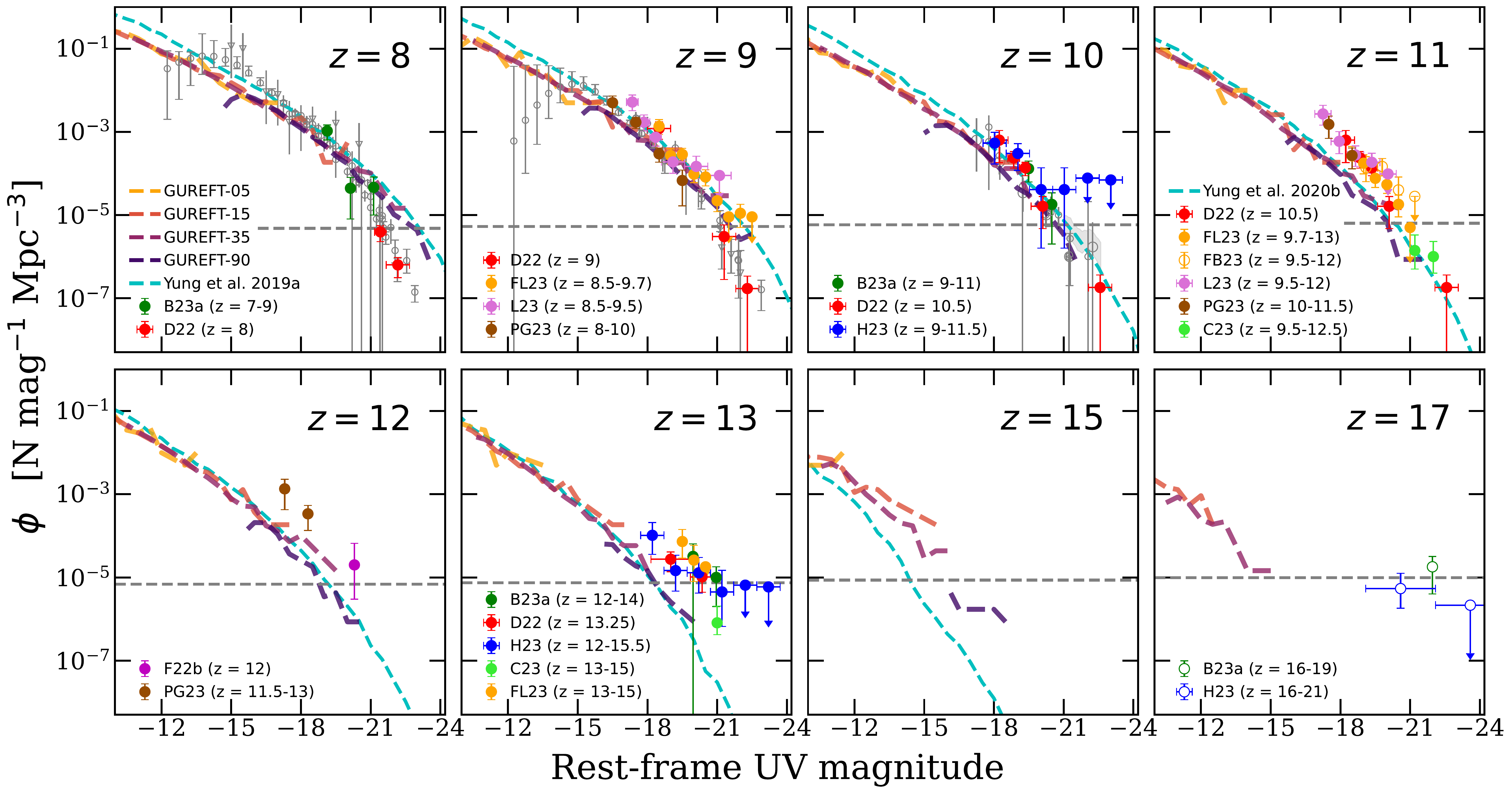}
    \caption{
        Predicted high- to ultrahigh-redshift rest-frame UV luminosity functions from the Santa Cruz SAM using \gureft\ merger trees. The UVLF based on merger trees extracted from the four different resolution \gureft\ boxes are shown with different colours, as indicated on the figure label. These results are compared to a compilation of observational measurements at high redshift (see text) and ultra-high redshift reported by \citet[][D22]{Donnan2022}, \citet[][H23]{Harikane2022}, \citet[][CEERS Epoch 1 only; hereafter \citetalias{Finkelstein2022b}]{Finkelstein2022b}, \citet[][full CEERS field; hereafter \citetalias{Finkelstein2023}]{Finkelstein2023}, \citet[][F22b]{Finkelstein2022a}, \citet[][B23a]{Bouwens2022b}, \citet[][PG23]{Perez-Gonzalez2023}, \citet[][L23]{Leung2023a}, and \citet[][C23]{Casey2023} as well as previous predictions from \citet[][$z \lesssim 10$]{Yung2019} and \citet[][$11 \lesssim z \lesssim 15$]{Yung2020a}. 
        The grey dashed horizontal line marks where one object is expected for a survey area similar to that of the full CEERS area ($\sim 100$ arcmin$^2$) with $\Delta z = 1$ centred around the redshift labelled for each panel.
        Measurements derived from past \textit{Hubble} and \textit{Spitzer} Space Telescopes based studies are shown in grey. See text for more details on the observational studies included in this comparison. 
        }
    \label{fig:gureft_UVLF_allz}
\end{figure*}

\begin{figure*}
    \includegraphics[width=2\columnwidth]{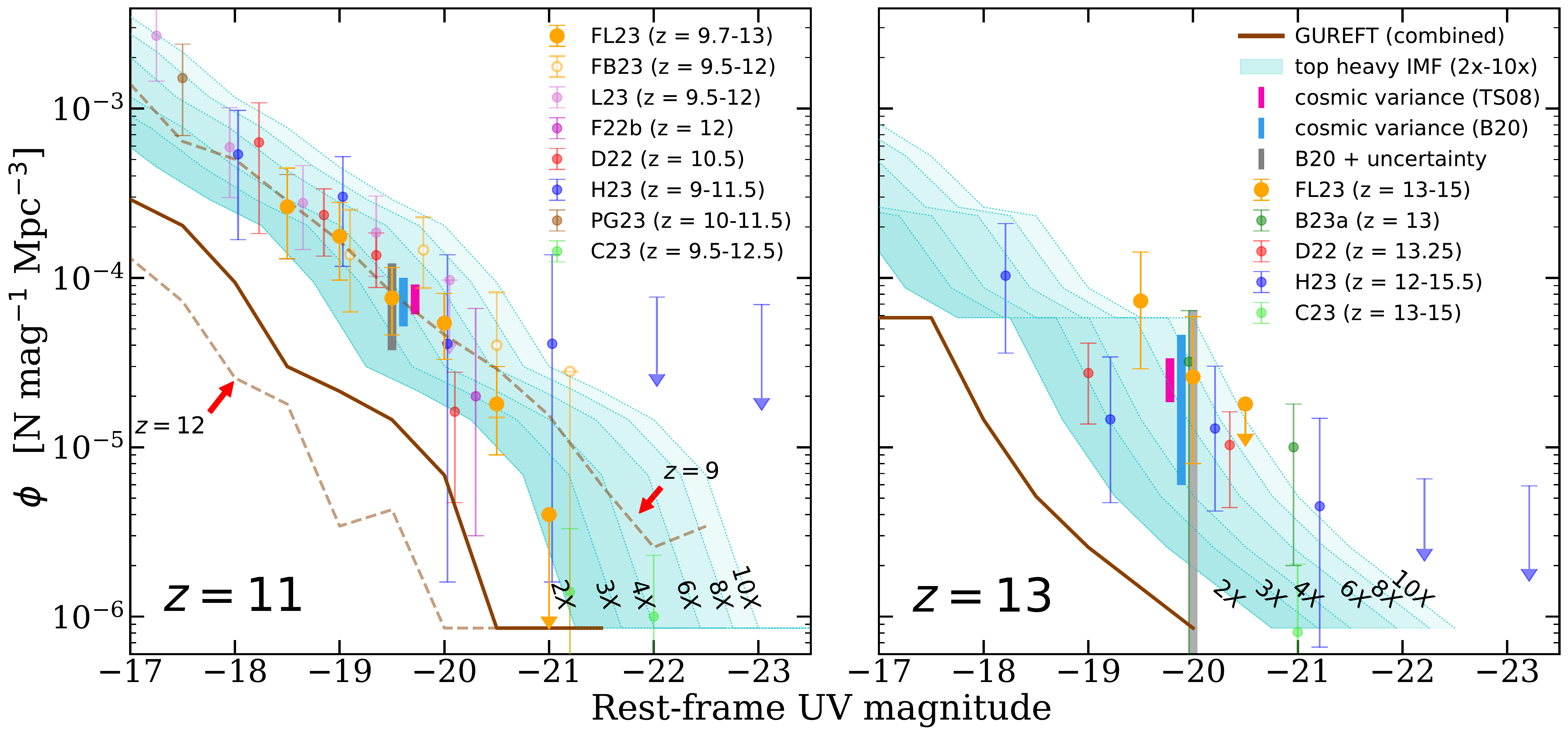}
    \caption{
        A more detailed look at the UV LFs at $z=11$ \textit{(left)} and $z=13$ \textit{(right)}. The combined outputs across the four \gureft\ boxes are shown by the brown line. In the left panel, in addition to the $z = 11$ predictions, we also show the predicted UV LFs at $z = 9$ and 12 with the light brown dashed line above and below, respectively, as the observational sample of F23 spans a redshift range of $z=9.5$--12.
        The observational measurements are the same as those shown in the corresponding panels in Fig.~\ref{fig:gureft_UVLF_allz}. In addition, we show two estimates of the 1$\sigma$ uncertainty on the number density due to cosmic variance with the blue and pink error bars (see text for a full description of how these were calculated). The grey bar shows a representative total error, including the quoted error bar from the observational study combined with the estimated cosmic variance in quadrature. The light blue shaded regions and lines show the effect of ``boosting'' the UV luminosity of all galaxies in our model by a factor of 2-10, as indicated on the plot label. This illustrates that a fairly modest boost of a factor of $\sim 4$ could bring our predictions into agreement with the observational measurements. 
    }
    \label{fig:gureft_UVLF_var}
\end{figure*}

\begin{figure*}
    \includegraphics[width=2\columnwidth]{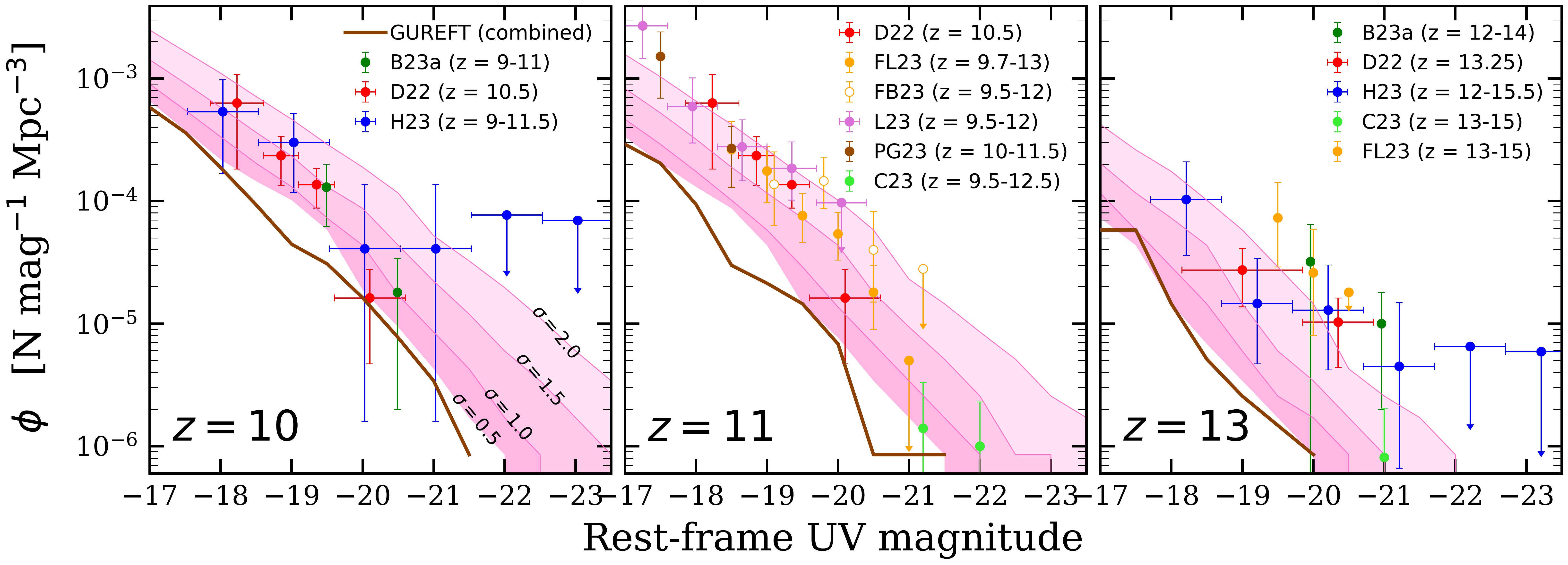}
    \caption{
        A more detailed look at the UV LFs at $z=10$ \textit{(left)}, $z=11$ \textit{(middle)}, and $z=13$ \textit{(right)}, where the fiducial predictions from this work made with the \gureft\ merger trees are shown in brown, observational compilations as shown in the corresponding panels in Fig.~\ref{fig:gureft_UVLF_allz}, and the potential effects of UV stochasticity shown with different shades of pink as labelled on the plot panels. This effect is implemented by adding a stochastic component to the predicted UV magnitudes of individual galaxies in post-processing, assuming the random component follows a Gaussian distribution centred at zero with standard deviation $\sigma$. We find that bursty star formation with a scatter of $\sigma \simeq 1.5$--2 is required to bring our fiducial model into agreement with observations at $z=11$--13.
    }
    \label{fig:gureft_UVLF_burst}
\end{figure*}

\subsection{Distribution functions}

One-point distribution functions are a frequently used summary statistic for characterizing and comparing galaxy populations from observations and simulations.
In Fig.~\ref{fig:gureft_SMF_allz}, we present the stellar mass functions (SMFs) for galaxies across the full mass range that we can resolve between $z=8$ to 17 by combining the results from the four \gureft\ volumes. One can see from the excellent agreement in the regions where the different boxes overlap that, as shown previously in \citet{Gabrielpillai2022}, the convergence is excellent (i.e. the SAM results are insensitive to the resolution of the merger trees as long as the merger history is well resolved). 
These results are compared to SMFs derived from past \textit{Hubble} and \textit{Spitzer} observations \citep{Song2016}.
The \textit{flattening} or \textit{turnover} seen at the low-mass end of \gureft-35 and \gureft-90 is caused by incompleteness due to the mass resolution limit of the underlying cosmological simulations. 
The \gureft-05 and \gureft-15 boxes are designed to resolve low-mass halos and their merger histories, and thus here the turnover reflects the halo mass limit where atomic cooling becomes inefficient (we do not include processes that can cool gas below $10^4$ K). This has been previously explored using the extended Press-Schechter (EPS)-based merger trees \citep[e.g.][]{Somerville1999a} and a grid of halo masses as presented in \citet{Yung2019a}, which is also shown in this comparison. As previously shown, our model predictions are in very good agreement with the observationally derived stellar mass functions at $z=8$. We refrain from showing stellar mass functions for higher redshifts or for any of the recent \textit{JWST} based studies, as the uncertainties on these stellar mass estimates are still extremely large. 

Fig.~\ref{fig:gureft_UVLF_allz} presents the rest-frame UV luminosity functions (UV LFs) between $z=8$ to 17 for galaxies simulated by the SC SAM.
These predictions are compared to the recent \textit{JWST} observations by \citet{Finkelstein2022a, Finkelstein2022b,Finkelstein2023} from CEERS \citep{Bagley2022a}, \citet{Leung2023a} from NGDEEP \citep{Bagley2023}, \citet{Casey2023} from COSMOS-Web \citep{Casey2022}, \citet{Donnan2022}, \citet{Harikane2022}, and \citet{Bouwens2022b}.
In addition, we include pre-JWST measurements from \citet{Bouwens2014a} and \citet{Finkelstein2015}, and a compilation of results from \citet{Finkelstein2016}. These include observations from lensed fields from \citet{Livermore2017} and \citet{Bouwens2022}, as well as the latest results derived from past \textit{Hubble} and \textit{Spitzer} Space Telescope observations presented by \citet{Oesch2018}, \citet{Stefanon2019}, \citet{Bowler2020}, and \citet{Finkelstein2021}. We show the $z\sim 17$ luminosity function estimates of \citet{Bouwens2022} and \citet{Harikane2022} as open points to illustrate how these published results compare with our model predictions. However, the single galaxy from the CEERS field \citep[first discovered by][]{Donnan2022} that went into the \citet{Bouwens2022} estimate has now been shown to be at $z\sim 5$ \citep{ArrabalHaro2023a}. The \citet{Harikane2022} estimate is based on two galaxies, the same CEERS galaxy now known to be at lower redshift, and a candidate from the Stephan's Quintet field (S5-z16-1). Similar to the CEERS object, S5-z16-1 has a bimodal photometric redshift solution with one peak at $z=16.4$ and another at $z=4.9$. The UV luminosity implied by the $z=16.4$ solution would be extremely high ($M_{\rm UV}=-21.6$). One can see that even one object of this luminosity at this redshift would be in very strong tension with our model predictions. 

In Appendix \ref{sec:tabulated}, we provide the tabulated data for the SMFs and UV LFs by combining the predicted galaxies across the four simulated volumes within the mass (and luminosity) range where the galaxy populations are complete. This approach is similar to the one adopted by \citet{Vogelsberger2020a} to produce \textit{JWST} predictions from the suite of IllustrisTNG simulations \citep[e.g.][]{Nelson2018, Pillepich2018, Nelson2019}. 

As also shown in previous works \citet{Yung2019a}, the rest-UV luminosity functions predicted by our models are in good agreement with observations at $z=8$--10. Note that the predictions shown here do not include any dust attenuation, so the small excesses in our predicted UV LFs at the bright end at $z\sim 8$ and 9 could be cured by invoking a moderate amount of extinction (see \citet{Yung2019a}, which did include modelling of dust attenuation). We also note that the free parameters used in this work are based on the re-calibration presented in \citet{Gabrielpillai2022}. In particular, the slope for the stellar feedback-mass outflow relation (see \citet{Yung2019} for full description) has been slightly increased from $\alpha_\text{rh} = 2.8$ to 3.0, resulting in higher mass outflow rates in low mass halos, and hence lower masses and luminosities for the galaxies hosted by those halos.
As a result, the faint-end (low mass-end) slopes for galaxies presented in this work are slightly shallower than those presented in previous works.
This is expected and is in agreement with the controlled experiment shown in \citet{Yung2019, Yung2019a}.
By $z=11$, we see a moderate discrepancy between our predicted UV LF and the observational measurements, with the number densities predicted by our models an order of magnitude or more lower than the observational estimates. The discrepancy is about a factor of 30 at $z=13$. 

The number density constraints estimated for observed galaxies presented in previous figures suffer from various uncertainties, including photometric errors, redshift errors, and field-to-field variance due to Poisson sampling statistics as well as galaxy clustering (field-to-field variance). Photometric errors can cause a so-called ``Eddington bias'', where the much more numerous intrinsically low-luminosity galaxies are ``scattered'' into the higher luminosity bins, causing a flattening of the high-luminosity slope of the luminosity function). The full uncertainty budget has not been fully accounted for in the published error bars. 
Similarly, the simulated outputs also carry uncertainties due to cosmic variance associated with our relatively small simulated volumes.

\begin{figure*}
    \includegraphics[width=2\columnwidth]{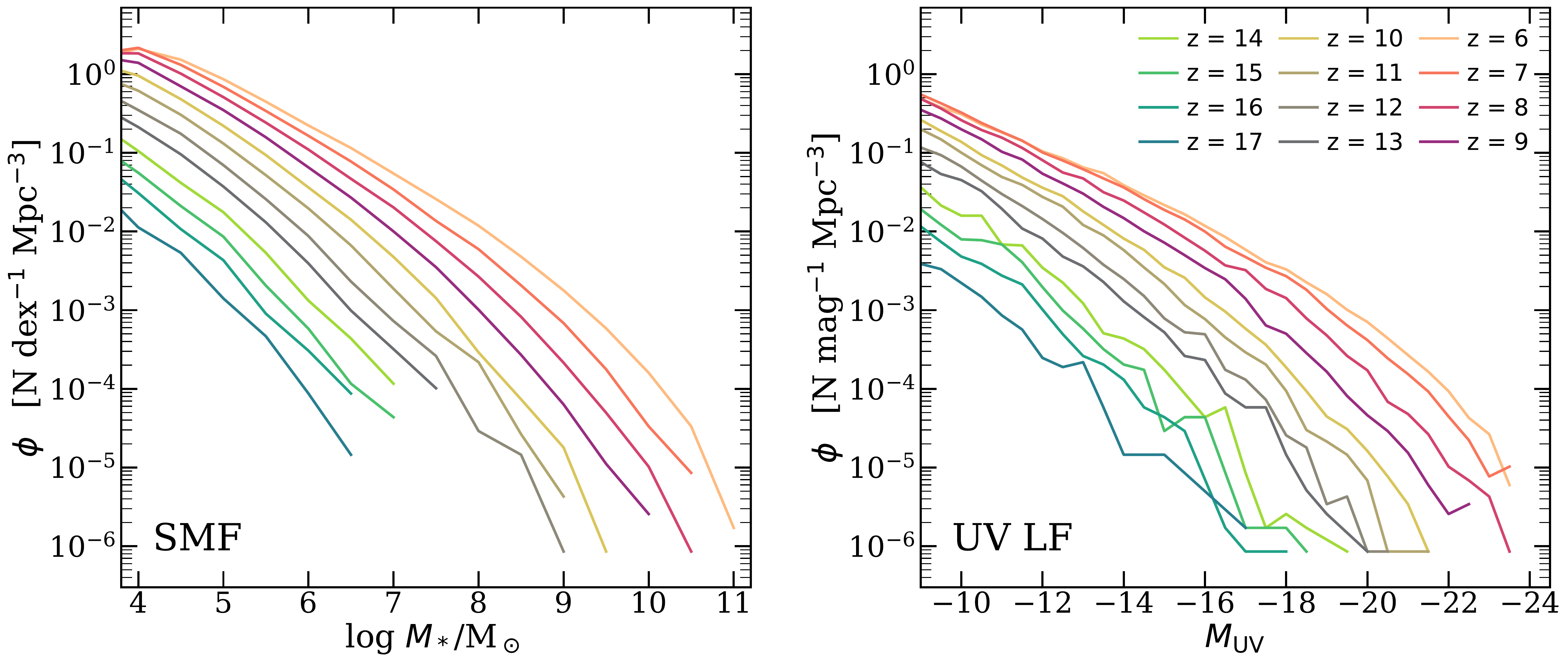}
    \caption{
        The evolution of the stellar mass function (left) and dust-free rest-frame UV luminosity function (right) over the redshift interval $z=6$ to $z=17$, using the combined \gureft\ volumes. Our models predict relatively strong evolution in both quantities. 
    }
    \label{fig:gureft_distFn_evo}
\end{figure*}

In order to better assess the level of tension between our theoretical predictions and the observational measurements, in Fig.~\ref{fig:gureft_UVLF_var} we show a ``zoomed in'' version of the $z=11$ and $z=13$ UV LF comparisons, where we focus on the luminosity range of the detected galaxy populations. Here we have selected the luminosity range where the simulated galaxies are best captured for each \gureft\ box and combine them. For instance, at $z = 12$ and 13, galaxies with $M_\text{UV} \leq -18$ are sourced from \gureft-90, and galaxies between $-18 \leq M_\text{UV} \leq -13$ are from \gureft-35.

Note that we do not attempt to correct for redshift uncertainties (which may shift the galaxies in redshift within these bins) or catastrophic redshift errors (which may cause galaxies that are at much lower redshifts to be counted in this ultra high-z population), as these corrections remain very uncertain and are not straightforward to model. We experimented with including photometric noise in our predicted luminosity functions, and find that it introduces a minor source of uncertainty in this comparison, so we do not show it here. 

We explore the potential impact on the number density of simulated galaxies due to cosmic variance. We show cosmic variance estimates from the cosmological simulation \textsc{bluetides} \citep{Feng2016, Bhowmick2020}, which is 400 h$^{-1}$ on a side, $\sim88$ and $\sim1492$ times larger than the box sizes of \gureft-90 and \gureft-35, respectively. We also show estimates from the cosmic variance calculator of \citet{Trenti2008}. A detailed discussion of how we obtained these estimates for each panel, and of uncertainties associated with them, can be found in Section~\ref{sec:discussion}. We have chosen the CEERS survey as a representative field for these estimates. 
We show a representative ``total error'' estimate as the sum in quadrature of the quoted error from \citetalias{Finkelstein2023} and the larger of our cosmic variance error estimates. 
In their $z\sim 11$ bin, the uncertainty due to Poisson sampling and photometric errors combined (a fractional error of $\sim $0.44 at $M_{\rm UV} =-20$) is comparable to that from cosmic variance (fractional error of $\sim 0.2-0.3$). At $z\sim 14$ and $M_{\rm UV} =-20$, the quoted fractional error is 100\%, and the uncertainty due to cosmic variance is $\sim 0.3-0.77$.

In Fig.~\ref{fig:gureft_UVLF_var}, we also show the impact of theoretical uncertainties in the stellar initial mass function. For our fiducial calculations, we have adopted an IMF that is similar to the standard \citet{Chabrier2003a} IMF which is typical for galaxies in the nearby Universe. However, there are multiple reasons to think that the IMF might be top heavy in the early Universe (as discussed in more detail in Section~\ref{sec:discussion}). It has been suggested that stellar populations could be 2-10 times brighter in the UV if the IMF is top heavy \citep[e.g.][]{Raiter2010}. The blue shaded regions in Fig.~\ref{fig:gureft_UVLF_var} show what happens to our predictions if we simply boost the luminosity of each galaxy in our simulation by a factor of 2--10. One can see that, at $z=11$--13, a fairly modest ``IMF boost'' of a factor of four would bring our model predictions into reasonable agreement with the observational estimates. 

Although our model includes merger-triggered bursts, there could be short timescale star formation stochasticity due to ISM processes that we do not resolve or attempt to represent in our models. Given there are fewer bright objects, adding stochasticity causes more galaxies to scatter into the bright end of the UVLF, in an effect similar to Eddington bias.  In order to explore the effect of UV stochasticity in a simple way, we add a stochastic component to the predicted UV magnitudes in post-processing. The stochastic component is assumed to follow a Gaussian distribution centred at zero for specified values of the standard deviation $\sigma_{\rm UV}$. 
We repeat the process of adding the stochastic component and computing UV LFs for 2,500 independent realizations and report the median of the outcome.
In Fig.~\ref{fig:gureft_UVLF_burst}, we show the effect of $\sigma_{\rm UV}=0.5$, 1.0, 1.5, and 2.0; and compare them to the same observational constraints shown in Fig.~\ref{fig:gureft_UVLF_allz}. We find that a relatively large value of $\sigma\simeq 2$ is required to match the observations at $z\sim 11$--13.

\begin{figure}
    \includegraphics[width=\columnwidth]{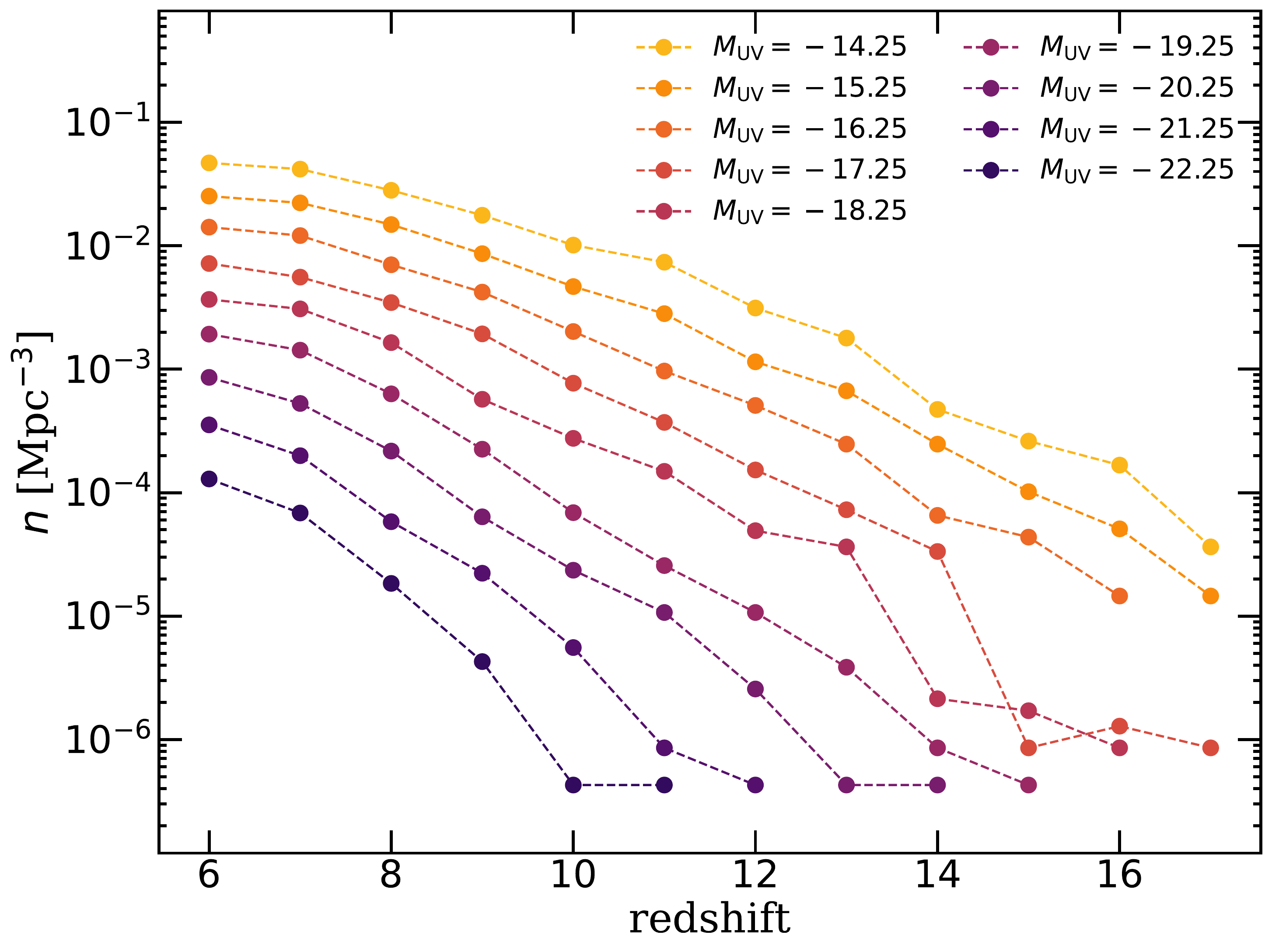}
    \caption{The predicted number density of galaxies in a fixed bin of $M_\text{UV}$ (bin width $\Delta M_\text{UV} = 1$) as a function of redshift. The number density of luminous galaxies is predicted to evolve more rapidly than that for lower luminosity galaxies.
    }
    \label{fig:gureft_density_binMuv_evo}
\end{figure}

In Fig.~\ref{fig:gureft_distFn_evo} we show how the composite stellar mass function and rest-UV luminosity function evolve from $z=6$ to 17. These are compiled by combining all of the \gureft\ volumes. This figure highlights the relatively rapid predicted evolution of both the SMF and UVLF. This appears to be in tension with the current observational results, which suggest weaker evolution over cosmic time at the highest redshifts. Fig.~\ref{fig:gureft_density_binMuv_evo} shows the same quantity but sliced in a different way -- this shows the number density in a bin of $M_\text{UV}$ as a function of redshift. The number density of luminous galaxies is predicted to evolve more rapidly than that of lower luminosity galaxies, as has also been predicted (and observed) at lower redshifts (sometimes called `downsizing'). 

\subsection{Scaling relations at ultra-high redshift}

Scaling relations are conditional relationships between two physical or observable properties of galaxies, and they are
important diagnostics of the physics that shapes galaxy properties.  In addition to rest-frame $M_\text{UV}$ and $M_*$, physical properties such as star formation rate (SFR) and UV spectral slope ($\beta_\text{UV}$) can be inferred from the measured multi-band photometry with SED fitting techniques \citep[e.g.][]{Johnson2021, Carnall2018, Iyer2019, Boquien2019}. In this subsection, we compare outputs from the SC SAM to scaling relation constraints inferred by observations. In these studies, physical properties such as stellar masses are obtained through SED fitting. It is important to keep in mind that these techniques rely on a complex modelling procedure, and the estimates can have large uncertainties due to corresponding uncertainties in the underlying simple stellar population models, and assumed priors regarding star formation history, chemical evolution, dust, etc. \citep[e.g.][]{Papovich2011, Tacchella2021}. 
Furthermore, these inferred properties are subject to uncertainties in distance.

\begin{figure}
    \includegraphics[width=\columnwidth]{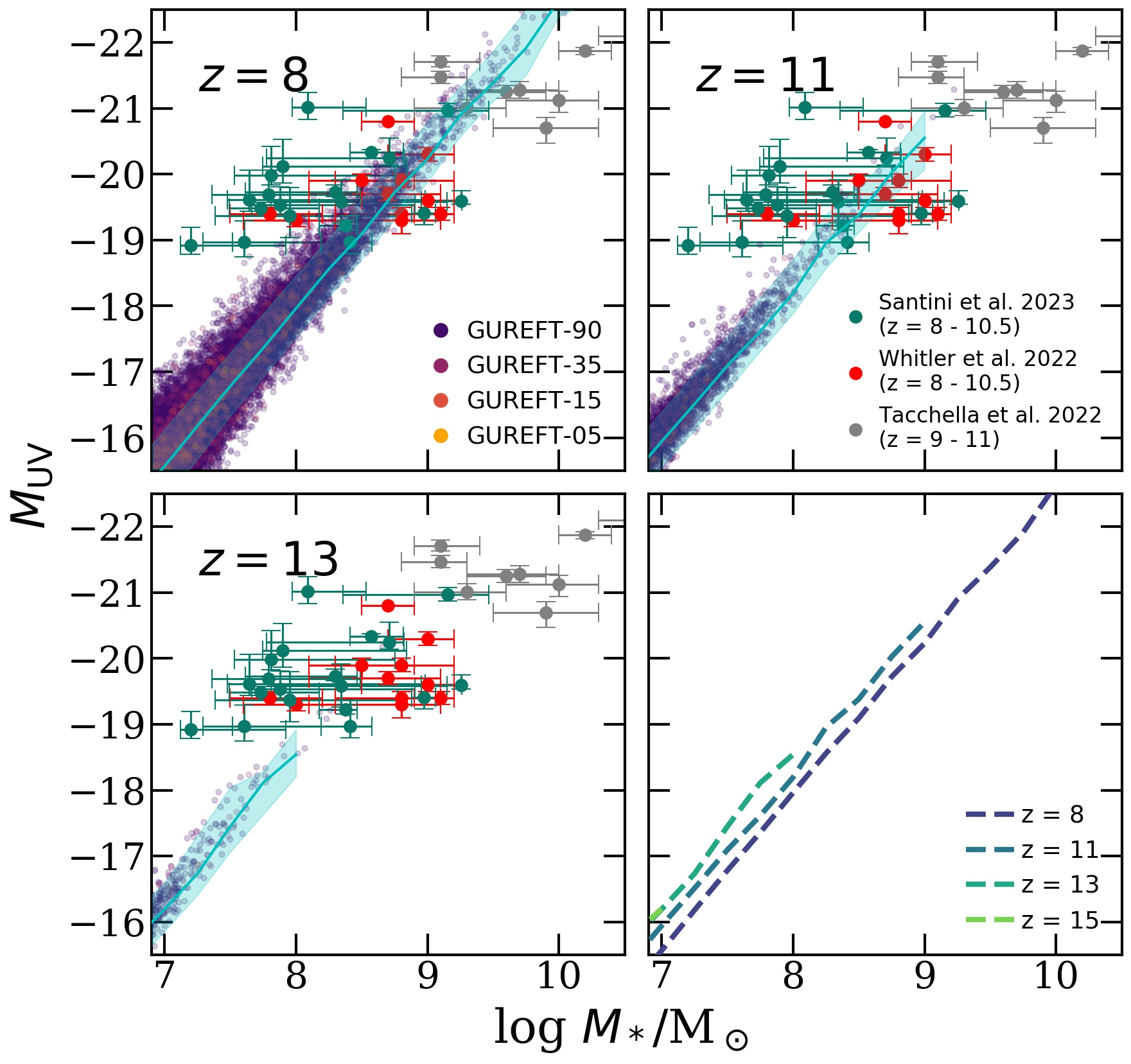}
    \caption{
        Scaling relation for stellar mass vs. rest-UV magnitude ($M_*$--$M_\text{UV}$ relation) for simulated galaxies at $z = 8$, 11, and 13. The small data points show SAM galaxies from the \gureft\ boxes. The cyan line marks the median and the shaded area marks the 16th and 84th percentiles for galaxies in bins of $M_*$. These results are compared with stellar mass estimates from SED fitting based on \textit{JWST} observations from \citet[][red]{Whitler2022} and \citet[][green]{Santini2023}, and \textit{Hubble} and \textit{Spitzer} observations from \citet[][grey]{Tacchella2021}. The last panel summarizes the evolution of $M_*$--$M_\text{UV}$ relation. 
    }
    \label{fig:4panel_mstar_muv}
\end{figure}

\begin{figure}
    \includegraphics[width=\columnwidth]{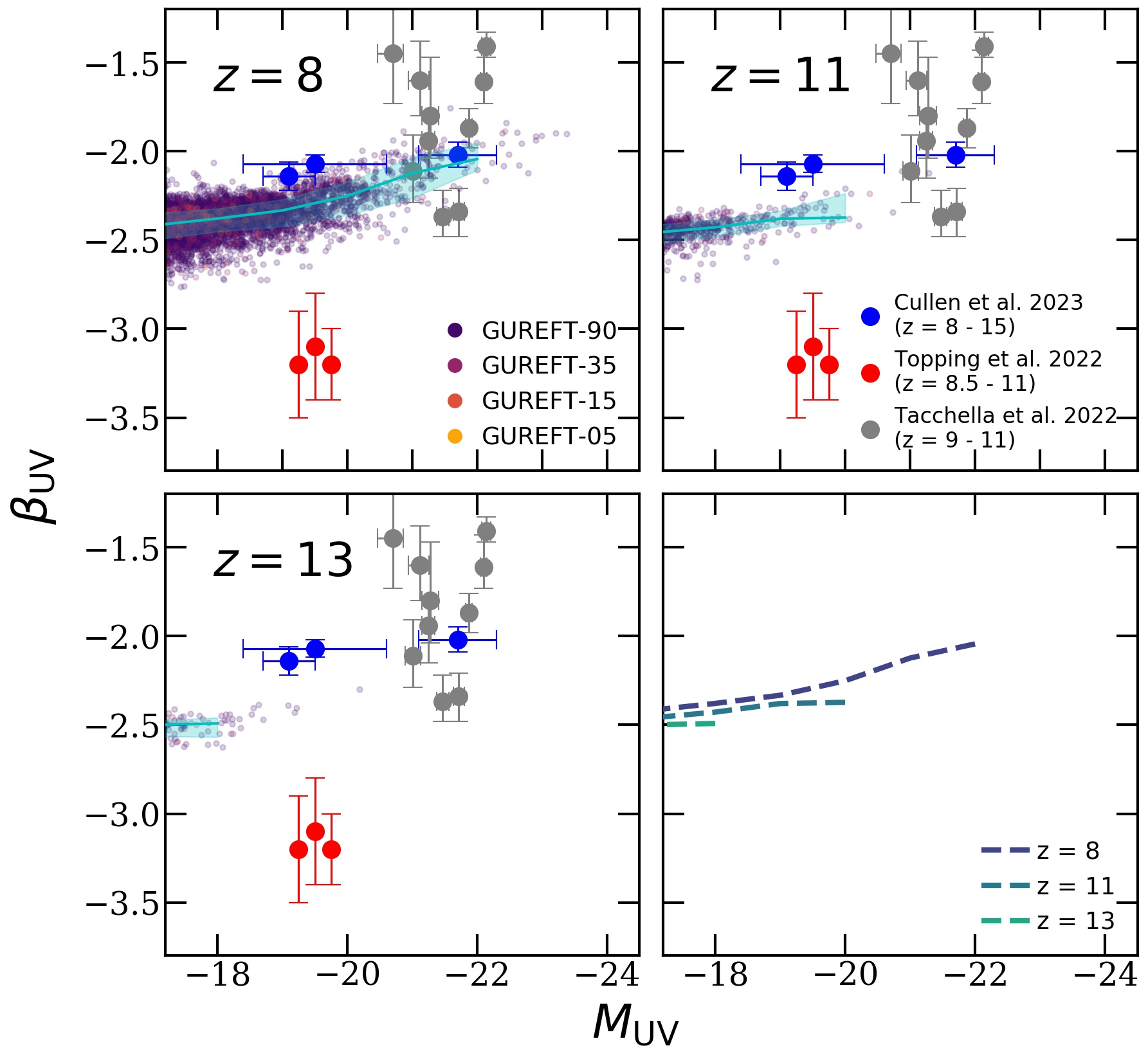}
    \caption{
        Rest-UV magnitude ($M_\text{UV}$) vs. rest-UV spectral slope ($\beta_\text{UV}$) relation for simulated galaxies at $z = 8$, 11, and 13. The small data points show SAM galaxies from the \gureft\ boxes. The cyan line marks the median and the shaded area marks the 16th and 84th percentiles for galaxies in bins of $M_\text{UV}$. These results are compared with measurements from 
        \textit{JWST} observations from \citet[][blue]{Cullen2022} and \citet[][red]{Topping2022}, and \textit{Hubble} and \textit{Spitzer} observations from \citet[][grey]{Tacchella2021}. The last panel summarizes the evolution of the $M_\text{UV}$--$\beta_\text{UV}$ relation.
    }
    \label{fig:4panel_muv_betauv}
\end{figure}

Fig.~\ref{fig:4panel_mstar_muv} shows the $M_*$--$M_\text{UV}$ relation from our predicted SAM galaxies, where individual galaxies are shown by the data points and the median and 16th and 84th percentiles are shown with the cyan line and shaded area.
These results are compared to results presented by \citet{Whitler2022} derived from NIRCam imaging in the CEERS field and the flexible Bayesian SED fitting code \textsc{Prospector} with a built-in `continuity' prior. 
In addition, we include constraints derived from past \textit{Hubble} and \textit{Spitzer} observations by \citet{Tacchella2021}. These quantities are also based on the \textsc{Prospector} tool, assuming a `bursty' version of a continuity prior. We note that while these galaxies span a wide range of redshift ($8 < z < 11$ for \citet{Whitler2022} and $9 < z < 11$ for \citet{Tacchella2022}) with relatively large uncertainties in photometric redshift, we show all observed galaxies across all panels in Fig.~\ref{fig:4panel_mstar_muv} for discrete redshift snapshots at $z = 8$, 11, and 13 from our simulations. The median of the $M_*$--$M_\text{UV}$ relation are overlaid in the last panel. It is intriguing that the model predictions for the relationship between stellar mass and rest-UV luminosity are in good agreement with the observational estimates at $z=8$ and $z=11$. In the $z=13$ bin, there are very few model galaxies in the luminosity range of the observations, but the observed galaxies seem to lie along a plausible extrapolation of the simulated relation for lower luminosity galaxies. Also notable is the rather weak evolution in the $M_*$--$M_\text{UV}$ over this redshift interval from $8 < z < 13$, with galaxies predicted to be slightly more luminous for a given stellar mass at higher redshift.

Fig.~\ref{fig:4panel_muv_betauv} shows a comparison between the modelled and observed $M_\text{UV}$ and rest-UV slope $\beta_\text{UV}$.
For the simulated galaxies, we compute $\beta_\text{UV} = -0.4(M_\text{FUV} - M_\text{NUV})/\log(\lambda_\text{FUV}/\lambda_\text{NUV}) - 2$ with magnitude in far UV (FUV, centred at 1600\,\AA) and near UV (NUV, centred at 2300\,\AA) bands. These results are compared to results from \textit{JWST} observations reported by \citet{Topping2022} and \citet{Cullen2022}. In addition, we also show results from \textit{Hubble} and \textit{Spitzer} observations reported by \citet{Tacchella2021}. The median of the $M_\text{UV}$--$\beta_\text{UV}$ relations from all three redshift snapshots are overlaid in the last panel. Some of the observed galaxies have significantly redder values of $\beta_\text{UV}$ than our models predict, which could be due to the presence of dust in some of these galaxies. Additionally, our models assume a metallicity floor of $10^{-3}$ Z$_{\rm sun}$, and the BPASS spectral synthesis models do not include Pop III type tracks or SEDs. The extremely blue slopes reported by \citet{Topping2022} are an exciting hint that we may be witnessing the formation of Pop III stars in some of these objects. The broad range of UV slopes seen in the observations suggests that the galaxy population could be a mixture of Pop II and Pop III stars at these epochs. 

\begin{figure}
    \includegraphics[width=\columnwidth]{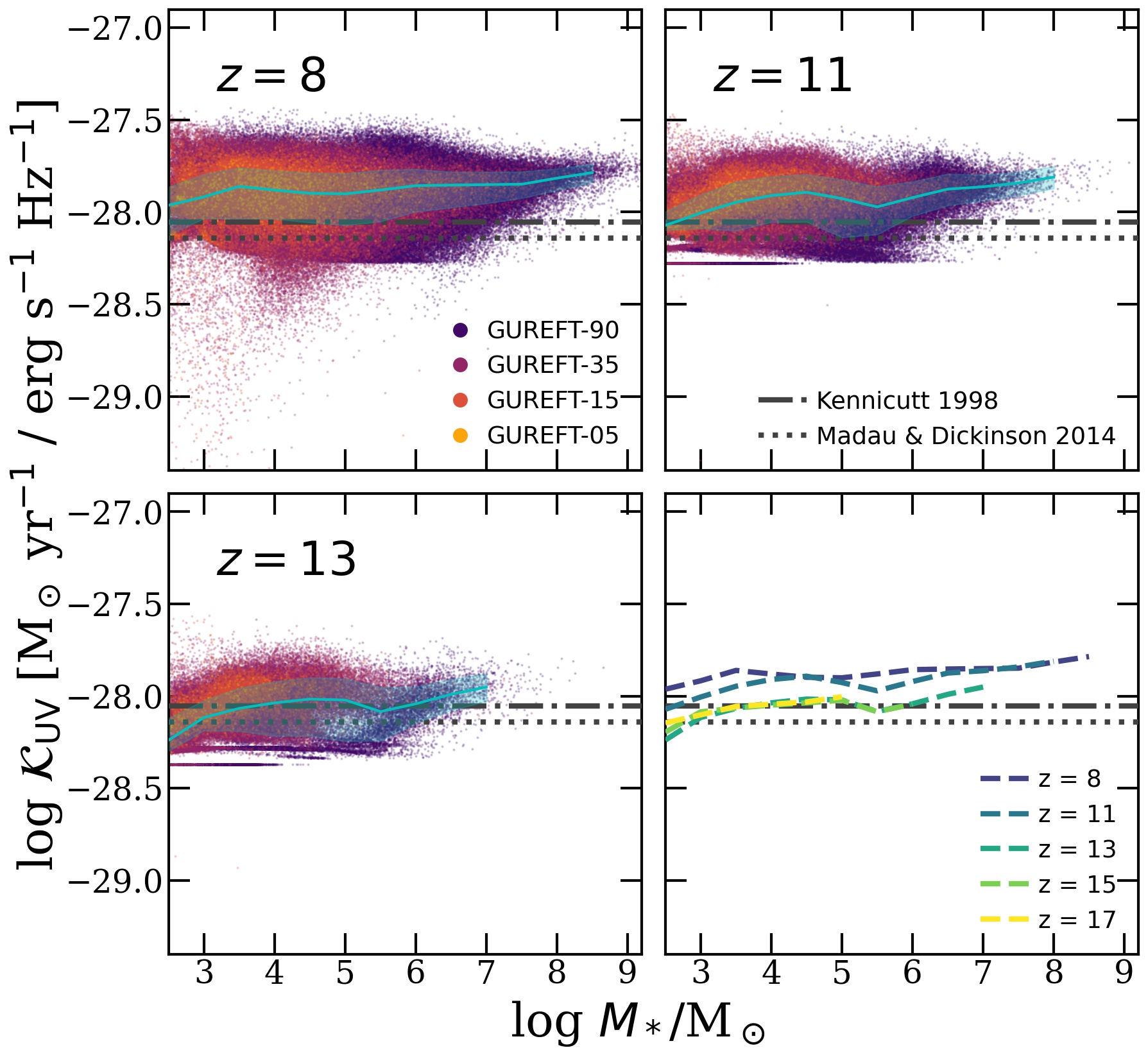}
    \caption{
        The ratio of star formation rate to rest-UV luminosity, ${\cal{K}}_\text{UV}$, as a function of stellar mass for simulated galaxies at $z = 8$, 11, and 13. The small data points show SAM galaxies from the \gureft\ boxes. The cyan line marks the median and the shaded area marks the 16th and 84th percentiles for galaxies in bins of $M_*$. These results are compared with nominal values adopted by past studies \citet[][grey dot-dashed line]{Kennicutt1998} and \citet[][grey dotted line]{Madau2014}. The last panel summarizes the evolution of this relation from $z=8$ to 17.
    }
    \label{fig:4panel_mstar_kuv}
\end{figure}

\begin{figure}
    \includegraphics[width=\columnwidth]{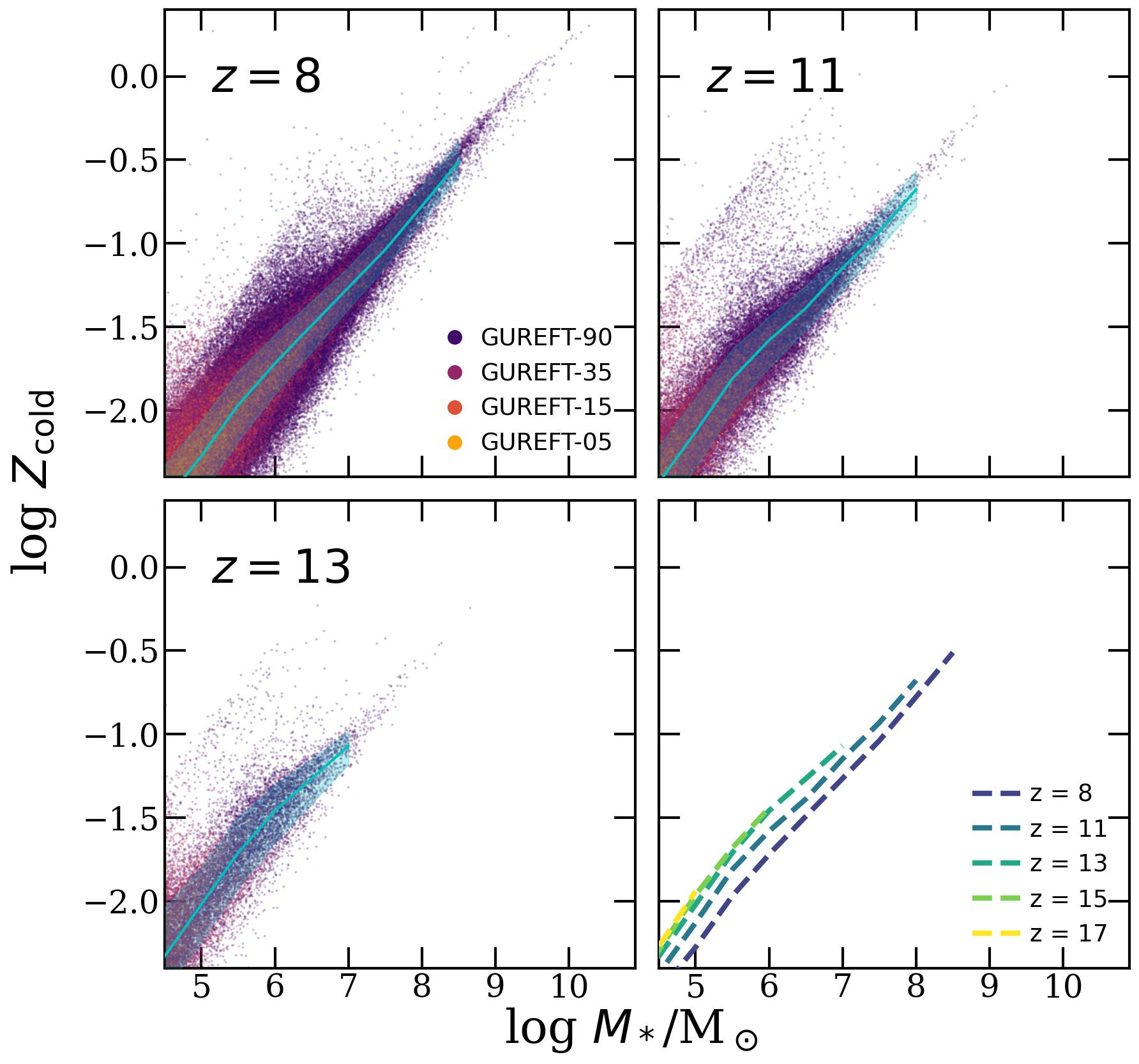}
    \caption{Metallicity of cold ISM gas as a function of stellar mass for simulated galaxies at $z = 8$, 11, and 13. The small data points show SAM galaxies from the \gureft\ boxes. The cyan line marks the median and the shaded area marks the 16th and 84th percentiles for galaxies in bins of $M_*$. The last panel summarizes the evolution of this relation from $z=8$ to 17.       
    }
    \label{fig:mass-metallicity}
\end{figure}

Fig.~\ref{fig:4panel_mstar_kuv} shows the relation between $M_*$ and $\cal{K}_\text{UV}$. The factor $\cal{K}_\text{UV}$ is an empirically measured conversion factor that relates the rest-frame UV luminosity to the star formation rate $\text{SFR} = {\cal{K}_\text{UV}} \times L_\nu(\text{UV)}$, where we consider the FUV luminosity and SFRs averaged over 100 Myr for this comparison.
Similar to previous figures, we show the individual galaxies with data points and the median and 84th and 16th percentiles with cyan lines and shaded regions.
We compare our simulations with values reported by \citet{Madau2014}, ${\cal{K}}_\text{UV} = 0.72\times10^{-28}$\Msun\ yr$^{-1}$ erg$^{-1}$ s Hz and ${\cal{K}}_\text{UV} = 0.88\times10^{-28}$\Msun\ yr$^{-1}$ erg$^{-1}$ s Hz as adopted by \citet{Kennicutt1998}. We note that these values are adjusted from a \citet{Salpeter1955} to \citet{Chabrier2003a} IMF by multiplying by a factor of 0.63 \citep{Madau2014}.
Since $\cal{K}_\text{UV}$ is effectively the efficiency of producing observed FUV light for a given SFR, it is expected that its value will be sensitive to star formation and chemical enrichment histories. 
The medians of the $M_*$--$\cal{K}_\text{UV}$ relations at $z=8$ to 17 are overlaid in the last panel.
As expected, galaxies in the early universe contain younger stellar populations and are more efficient at producing UV light. 
Similarly, Fig.~\ref{fig:mass-metallicity} shows the relation between metallicity of cold ISM gas ($Z_\text{cold}$) and $M_*$.

Additional scaling relations with dark matter halo mass are presented in Appendix \ref{sec:halo_connection}.

\section{Discussion}
\label{sec:discussion}

In this section, we discuss the implications of recent \textit{JWST} observations in the context of theory and simulations. We also discuss the caveats and uncertainties in the results presented in this work.

\subsection{Dark matter halos at ultra-high redshift}

Halo assembly and merger histories form the backbone of galaxy formation simulations, especially for the semi-analytic modelling approach. Cosmological simulations are subject to tension between simulated volume and mass resolution, subject to available computational resources. While extremely large volume is required to capture overdense environments where the very rare, massive halos are found, it is also critical to have sufficient mass resolution to properly resolve halos' assembly histories. While this tension is always present in modelling galaxy formation (since we can observe a much larger dynamic range of galaxy environments than we can currently simulate), there has been less attention directed at creating a suitable suite of N-body simulations for studying structure formation in the ``ultra-high redshift'' Universe ($z>6$) --- up until now. 
Establishing connections between halos identified across multiple simulation snapshots and constructing the full merger histories of halos also requires high temporal resolution. 
However,  most state-of-the-art cosmological N-body simulations store a limited number of snapshots at ultrahigh redshift (likely a thoughtful decision made to optimize storage usage), and these are insufficient to reconstruct merger histories of these halos.
For these reasons, many publicly available cosmological simulations do not deliver halo merger trees that can be used by SAMs to explore predictions for galaxies in the exciting redshift frontier that has recently been opened up by \textit{JWST} ($10<z<17$). We designed the \gureft\ simulation suite  specifically to overcome these barriers. In a companion paper \citep{Yung2023b}, we will present updated fitting functions describing the halo mass function and other useful halo property distributions in the ultra-high redshift universe from the \gureft\ suite along with a comparison with previous halo mass functions that have been used in the literature.

We add that the mass range probed by the higher resolution boxes, \gureft-05 and \gureft-15, may be probing a regime that can be affected by uncertainties in the physical nature of dark matter. In this work, we assumed `vanilla' $\Lambda$CDM. If the dark matter has a wavelike or ``fuzzy'' nature, this could substantially impact the low-mass end slope of the HMF \citep[][see \citealt{Hui2021} for a review]{Menci2017,Hui2017,Kulkarni2022}.
Additionally, if the dark matter is warm or self-interacting, this could also impact the abundances of low-mass halos at early times and their internal structures and formation histories \citep[e.g.][]{Menci2016,Lovell2018,Adhikari2022}.
The nature of dark energy could also impact structure formation at early times. Models with Early or Dynamical Dark Energy can predict higher abundances of low-mass halos at early times \citep{Menci2020,Klypin2020,Menci2022}.

\subsection{Comparison with other model predictions}
A comprehensive comparison between our predictions and those of the large number of other published models and simulations is beyond the scope of this work, but we comment briefly here on the general theory landscape and on a few key comparison points. First, we comment on the predictions of the ``Semi-analytic Forecasts for \textit{JWST}'' paper series \citep{Yung2019,Yung2019a,Yung2020}, which adopted the same semi-analytic model presented here with slightly different parameters, but implemented within merger trees constructed using the Extended Press-Schechter formalism. We show in Fig.~\ref{fig:gureft_SMF_allz} and Fig.~\ref{fig:gureft_UVLF_allz} that there is very good agreement between the Yung et al. predictions and those of the present work from $z\sim 8$--13, with the small differences on the faint end of the UVLF being largely due to the re-calibration of the stellar feedback parameters as discussed in Section~\ref{sec:results}. At $z>13$, our new and more robust calculations based on our new \gureft\ suite predict number densities that are higher by an order of magnitude or more than the EPS based calculations. We note that this has little to no impact on any of our published results or conclusions, as we deliberately focussed on redshifts $z \lesssim 12$ specifically because we did not trust the EPS merger trees at higher redshifts. 

Comparisons with other model predictions are shown in \citet{Harikane2022} for $z=12$ and $z=14$--16 (see their Figure 16) and \citetalias{Finkelstein2022b} (see their Figure~14) for $z\sim 8$--14. The general impression from these comparisons is that our predicted number densities at $z\sim 8$--10 are generally consistent (within a factor of a few) with those of large cosmological hydrodynamic simulations such as SIMBA \citep{Wu2020}, Millennium-TNG \citep{Kannan2022}, and THESAN (\citealt{Kannan2021}; see also \citet{Yung2019a} for a detailed comparison of our model predictions with simulation predictions available at that time out to $z=10$). The hydrodynamic simulation FLARES and the semi-analytic model DELPHI predict number densities at $m<28.5$ and $z\sim 11$ that are higher than ours by about an order of magnitude and a factor of 3, respectively (F23). Our predicted number densities are also very similar to those of the semi-empirical model {\sc UniverseMachine} \citep[see][for a detailed comparison]{Behroozi2020}. In summary, the significant under-prediction of $z\gtrsim 10$ galaxies observed by \textit{JWST} in physics-based models is a fairly generic result and certainly not specific to our models. 

\subsection{Observational uncertainties}

\subsubsection{Selection robustness, photometry, redshifts, completeness}
The recent observations with \textit{JWST} are providing confirmation of the measurements with Spitzer and HST probing out to $z \sim 9$--10, and extending the detection of galaxies into a new redshift frontier at $z \sim 10$--17. It is encouraging that the new measurements of UV LF at $z\sim 8$--10 with \textit{JWST} are highly consistent with the previously published estimates. Although the detection of galaxy candidates at ``ultra-high' redshifts $z\gtrsim 10$ with \textit{JWST} is very exciting, it is important to keep a few caveats in mind. First, many of these objects are \emph{candidates} that have been selected via a photometric redshift estimation procedure. Some studies first apply a ``Lyman break'' selection criterion (a colour-colour selection), and then compute photometric redshifts only for the colour selected high redshift candidates. Other studies (e.g. \citetalias{Finkelstein2022b} and \citetalias{Finkelstein2023}) compute photometric redshifts for all of the galaxies in the sample. It is not uncommon for galaxies to have a broad or bimodal redshift posterior. In the published studies, galaxies are assigned to the redshift bin corresponding to the ``best fit'' redshift. However, the estimates of rest-frame UV luminosity are subject to the distance uncertainties inherent in the redshift uncertainties. Strong emission line features are another source of uncertainty. The presence of strong emission lines in galaxy spectra can affect the redshift estimates \citep[e.g.][]{Finkelstein2022a} and are not always available in SED fitting templates \citep[e.g.][]{Larson2022}.

Some of the ``ultra high-z'' candidates may turn out to be at much lower redshifts. For example, multiple studies \citep{Naidu2022,Donnan2022,Finkelstein2022b,Bouwens2022a} presented a candidate in the CEERS field with a formal best-fit redshift of $16.4$, which has now been confirmed via NIRSpec spectroscopy to instead be at $z=4.9$. \citet{Harikane2022} have also reported a very luminous object with a redshift estimate of $16.41$ in the Stephan's Quintet field. Follow-up spectroscopy and millimeter interferometry will gradually confirm the true redshifts of this and other such objects that may be discovered in the near future through \textit{JWST} imaging.

It is also important to keep in mind that the areas and volumes of the surveyed fields to date are relatively small, as are the numbers of objects that go into luminosity function estimates. The updated CEERS $z\sim 11$ sample presented in \citetalias{Finkelstein2023} contains 27 galaxies. The $z\sim 13$ sample is even smaller. There are three (seven) candidates in the $13 < z < 15$ ($12 < z < 15$) redshift range in the full CEERS field reported in the analysis of \citetalias{Finkelstein2023}.
%There is only one reported candidate in the $12 < z < 14$ redshift range in the CEERS sample, which is assigned a redshift of $13.36^{+0.84}_{-1.08}$ by F23 and one of $11.6^{0.4}_{-0.5}$ by B23a. 
B23a report 4 total (2 robust) candidates in the $12 < z < 14$ redshift range in the SMACS0723 field and 3 total (2 robust) candidates in the GLASS field. Both of these fields are lensing cluster fields, and although B23a assume that no magnification correction is needed, it is curious that the angular number density in these fields is \emph{almost an order of magnitude higher} than that in the much larger (and presumably unlensed) CEERS field (see Table~\ref{tab:numbercounts}). This hints that there may be unaccounted for magnification or other systematic effects impacting the LF estimates from these fields. 

\begin{table}
	\centering
	\caption{Observed angular number density (number per arcmin$^{2}$ on the sky) of galaxies per field at $12 < z < 14$. }
	\label{tab:numbercounts}
	\begin{tabular}{lccc}
 \hline
 field & Area & $n$ (total) & $n$ (robust)\\
 \hline
CEERS Epoch 1 (\citetalias{Finkelstein2022b}) & 33.8 & 0.059  & 0.030 \\
CEERS complete (\citetalias{Finkelstein2023}) & 100 & 0.07 & 0.07 \\
 SMACS0723 (B23a) & 9.5 & 0.53  & 0.21 \\
 GLASS (B23a) & 7.0 & 0.43 & 0.30\\
 \hline
	\end{tabular}
 \end{table}

\citetalias{Finkelstein2022b}, \citetalias{Finkelstein2023}, and B23a provide extensive discussions of the overlap between high redshift galaxy candidates selected by different studies to date. B23a classify the fraction of candidates in different redshift and luminosity bins that are ``robust'', ``solid'', and ``possible'', finding that the estimated number density at $z=13$ differs by 1.5 orders of magnitude between the ``solid'' and the ``robust'' samples. 
\citetalias{Finkelstein2023} show a comparison of the photometric redshift estimates from different studies, which overall is fairly encouraging, but only includes objects in the CEERS field up to $z\sim 11.5$. \citetalias{Finkelstein2023} additionally report that the estimated rest-frame UV magnitudes for the same object can have significant scatter from one study to another (by typically a few tenths of a magnitude, but as much as a magnitude), and some studies can show a systematic offset in the estimated magnitudes by up to 0.3 magnitudes. All of the published luminosity functions have been corrected for completeness, and completeness corrections can also vary from one study to another even for the same raw observational data. However, the published results are typically presented only where the completeness is better than $\sim 50\%$, so this is not expected to be a dominant source of error at present.

There are other observational studies in the literature presenting ultra-high redshift galaxy candidates and luminosity function measurements from early \textit{JWST} observations \citep[e.g.][]{Rojas-Ruiz2020, Atek2022,Adams2022,Bagley2022,Naidu2022,Castellano2022a,Bouwens2022a,Labbe2022,Harikane2021a}. We omit these observations from our comparison, as they are typically based on smaller fields and in some cases were published before the updated instrument calibration. The \citetalias{Finkelstein2022b}, \citetalias{Finkelstein2023}, and B23a studies provide a detailed discussion of the consistency of their measurements with these earlier works. 

While we neglect dust throughout this work, in a scenario where dust enrichment happened quickly, the intrinsic UV luminosities will need to be even more boosted that shown in Fig.~\ref{fig:gureft_UVLF_var} to compensate for the attenuation effect. However, \citetalias{Finkelstein2023} and \citet{Morales2023} show that the colours remain fairly blue throughout the $z\sim9$ to 15, disfavouring the scenarios raised by the \citet{Ferrara2022}  and \citet{Mirocha2022} models.

\subsubsection{Cosmic Variance}
The fields that have been observed by \textit{JWST} to date are relatively small, and therefore it is expected that field-to-field variation in number density due to large scale structure (galaxy clustering) could be a significant (perhaps even dominant) source of uncertainty. In this work we follow a fairly common convention in separating the variance in the counts along different lines of sight into that due to Poisson point sampling and that due to galaxy clustering. We refer to the latter component alone as ``cosmic variance''. As is generally the case when a new discovery space for galaxy populations is opened up, our estimates of the magnitude of the cosmic variance for these populations is extremely uncertain, because we do not know how strongly these objects are clustered on larger scales. Due to these uncertainties, in this work, we used two different bracketing approaches to estimate the cosmic variance for the galaxy populations with $M_{\rm UV} \simeq -20$ at $z\sim 11$ and $z\sim 14$.   

The first method relies on abundance matching. Here, we assume that each halo above a critical mass is occupied by one galaxy, and we ask what is the minimum halo mass that would yield the observed galaxy number density at a given redshift. The clustering of these halos relative to the underlying dark matter density field provides an estimate of the ``bias'' $b$ for the observed galaxies, which then allows us to compute the cosmic variance $\sigma_{\rm cv}$ for a given field area, geometry, and redshift extent ($\sigma^2_{\rm cv} = b^2 \sigma^2_{\rm DM}$, where $\sigma_{\rm DM}$ is the variance of the dark matter density field) . We use the online calculator of \citet[][hereafter TS08]{Trenti2008} for these calculations\footnote{https://www.ph.unimelb.edu.au/~mtrenti/cvc/CosmicVariance.html}. We note that this tool has the limitation that the adopted cosmology is slightly different from the Planck-compatible cosmology that we adopt in the rest of this work, and that the estimates of halo number density and bias rely on the \citet{Sheth1999} approximation, which is known to be inaccurate at these extreme redshifts \citep{Yung2023b}. 

The second method is based on hydrodynamic simulations of galaxy formation. Specifically, we make use of the COSMIC\_VARIANCE\_AT\_HIGHZ calculator of \citet[][B20]{Bhowmick2020}, which is based on the predictions of the \textsc{bluetides} cosmological hydrodynamic simulations\footnote{https://github.com/akbhowmi/CV\_AT\_COSMIC\_DAWN}. However, these predictions rely on \textsc{bluetides} correctly predicting the clustering of the ultra-high-z galaxy population. Given that the UV luminosity functions predicted by \textsc{bluetides} at $z\gtrsim 11$ are about an order of magnitude lower than the nominal observational measurements, it is not clear how reliable these predictions are. However, the TS08 abundance matching approach and the B20 simulation based approach should provide a reasonable bracket on the cosmic variance estimates. 

\begin{table}
	\centering
	\caption{Cosmic variance estimates for representative magnitude and redshift bins. The area is given in arcmin$^2$. $N_g$ is the (completeness corrected) number of galaxies in the sample. $b_{\rm ave, TS08}$ is the average bias output by the TS08 calculator. mag is the magnitude bin used in the B20 calculator. $\sigma_{\rm cv, TS08}$ is the fractional root variance estimated using the TS08 method, and $\sigma_{\rm cv, B20}$ is the same quantity but estimated using the B20 method. }
	\label{tab:cosmicvariance}
	\begin{tabular}{lccccccc}
 \hline
 $\bar{z}$ & $\Delta z$ & Area & $N_g$ & $b_{\rm ave, TS08}$ & mag  & $\sigma_{\rm cv, TS08}$ & $\sigma_{\rm cv, B20}$ \\
 \hline
 11 & 3.3 & 100 & 16.5 & 12.3 & 28.5 & 0.20 & 0.32\\
 14 & 2.0 & 100 &  2.2 & 15.2 & 28 & 0.29 & 0.77 \\
 \hline
	\end{tabular}
 \end{table}

We provide estimates of the cosmic variance $\sigma_{\rm cv}$ using these two methods for representative redshifts ($z=11$ and 14) and magnitudes in Table~\ref{tab:cosmicvariance}. We note that $\sigma_{\rm cv}$ is the root fractional variance as defined in Eqn.~1 of \citet{Somerville2004}. As an illustrative example, we adopt the CEERS field with the completeness corrected numbers of galaxies in redshift bins $9.7<z<13$ and $13<z<15$ from \citetalias{Finkelstein2023}. The uncertainty due to cosmic variance for smaller fields such as SMACS0723 (9.5 arcmin$^2$) and GLASS (7.0 arcmin$^2$) are somewhat larger (e.g. $\sigma_{\rm cv}=0.29$ at $9<z<11$ and $\sigma_{\rm cv}=0.36$ for $12<z<14$ using the TS08 method). For simplicity, we assume that the footprints of the fields on the sky are  square (i.e. have an aspect ratio of unity). While elongated fields with aspect ratios much smaller than unity do have reduced cosmic variance relative to a square field of the same area \citep{Moster2011}, we have verified that for the field geometries investigated here, the cosmic variance is reduced by only 3\% for the $z=11$ field and 1.4\% for the $z=14$ field if we adopt the actual aspect ratio. 

As anticipated, the $\sigma_{\rm cv}$ estimates from the B20 method are significantly larger than those from TS08. This is because the implied host halo masses and therefore halo and galaxy biases in these two pictures are quite different. The average biases in the abundance matching picture, as obtained from the TS08 calculator, are given in Table~\ref{tab:cosmicvariance} and are $b \simeq 12$--15. As we noted, B20 predicts a lower galaxy number density at these redshifts than the observations (but their predicted UV LFs at $z=11$ and $z=13-14$ are quite consistent with our SAM predictions). Taking the observational LF estimates at face value, the different number densities could be reconciled either by assuming that the galaxies occupy the same mass halos but are brighter (as in our boosted ``top heavy IMF'' model), or by assuming that galaxies of a given luminosity actually occupy lower mass halos, which are more abundant. The abundance matching approach effectively puts galaxies in the ``right'' mass halos to reproduce the observed number density. These are lower in mass, and less clustered, than what \textsc{bluetides} predicts. 

\subsection{What have we learned about the physics of galaxy formation?}

With robust halo catalogs and merger trees spanning halos over a mass range of $\sim 10^{5.5}$--{$10^{12.5}$} from $z\sim20$--6, we are able to compute predictions for the properties of ultra-high redshift galaxies over an unprecedented dynamic range. Our predictions are based on the well-established Santa Cruz semi-analytic models, which adopt a very standard set of physical processes, including atomic cooling, partitioning of galaxies' ISM into multiple phases, star formation and stellar feedback, chemical evolution, and black hole seeding, growth, and feedback. Although these models contain simplified representations of these physical processes, we have shown in the past that the predictions for many key galaxy properties are remarkably similar to those from more detailed (and much more computationally expensive) numerical hydrodynamic simulations \citep{Yung2019, Gabrielpillai2022, Gabrielpillai2023}. 

One of the puzzles of galaxy formation manifested in observations of nearby galaxies is why star formation is so inefficient and why galaxies and halos capture such a small fraction of available baryons \citep[e.g.][]{Moster2010, Behroozi2013a}. The standard solution to this puzzle is that heating and winds from massive stars and supernovae eject baryons and make star formation inefficient in low mass halos, and similar processes associated with accreting black holes eject gas and prevent it from cooling in massive halos \citep[e.g.][]{Somerville2015a}. Our models, like all large volume cosmological simulations, contain parameterized phenomenological treatments of these key ``feedback'' processes, as it is infeasible to simulate the full range of scales and physical processes. These parameters have been tuned by hand to reproduce key global observable (or semi-observable) properties of galaxies in the nearby Universe, such as their stellar-to-halo mass relation, cold gas fractions, stellar metallicities, etc. There is absolutely no guarantee that the \emph{parameterizations} of key physical processes such as stellar feedback scale correctly in terms of the physical properties that are actually relevant. Therefore it is actually a bit surprising that the models perform as well as they do out to such high redshifts ($z\sim 10$). In this section, we ask what we learn from the apparently growing tension between the model predictions and the observational estimates at $z \gtrsim 10$.

We note that we have verified in the past that under our current implementation, black hole feedback has a negligible impact on the galaxy populations at these very high redshifts, so the main processes shaping galaxy properties in our models are the rate of gas accretion and cooling, the efficiency with which cold gas can form stars, and the efficiency of stellar driven outflows, which eject gas from the ISM and suppress future star formation. Taking the observational estimates at face value for purpose of this discussion, we consider four hypotheses for why our models might be predicting lower number densities of galaxies than the observations suggest: 
\begin{enumerate}
    \item The standard $\Lambda$CDM model does not predict early enough formation of halos massive enough to host the observed galaxies.
    \item Cooling is inefficient
    \item Star formation is inefficient
    \item Stellar feedback is too strong
\end{enumerate}

\begin{figure*}
    \includegraphics[width=2\columnwidth]{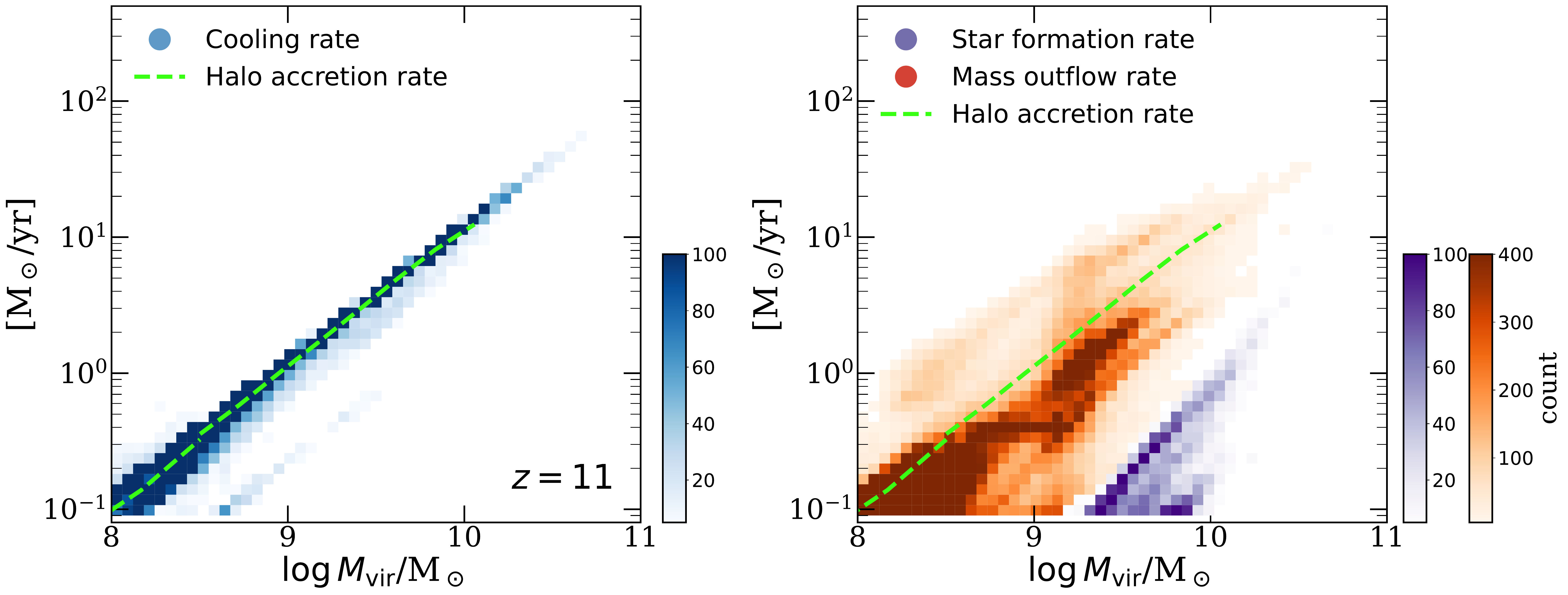}
    \caption{Diagnostics of several key physical processes in the Santa Cruz SAM at $z=11$. The green line in the leftmost panel shows the maximum accretion rate of baryons into dark matter halos, $f_{\rm b} \dot{M}_{\rm halo}$, where $f_b$ is the universal baryon fraction and $\dot{M}_{\rm halo}$ is the total halo mass accretion rate averaged over 1 Gyr. The blue points show the cooling rate of gas in the Santa Cruz SAM, illustrating that cooling is very efficient. The middle panel shows the SFR (blue) and the mass outflow rate from the ISM (brown), illustrating that the mass loading factor in the SAM is quite high, such that a gas is ejected from the ISM at a rate of 2 to 200 times the SFR. 
    }
    \label{fig:coolingfb}
\end{figure*}

\begin{figure}
    \centering
    \includegraphics[width=\columnwidth]{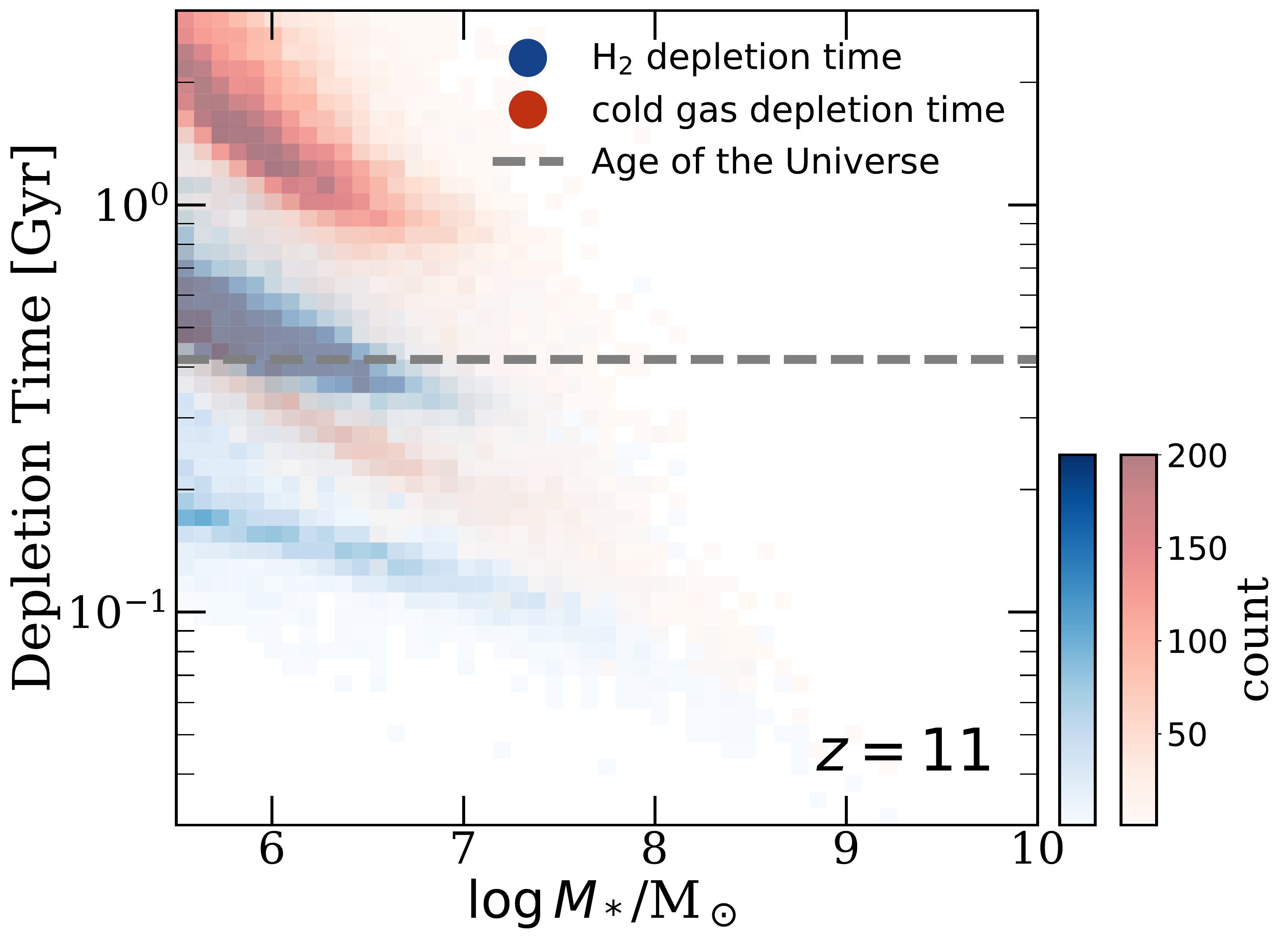}
    \caption{This shows the gas depletion time, which is given by the gas mass divided by the SFR, for molecular gas (orange) and for the total cold ISM gas (atomic, molecular, and ionized; turquoise) at $z = 11$. The horizontal grey line shows the corresponding age of the Universe. Our models actually predict that the majority of galaxies at $z\sim11$, especially the more massive ones, have depletion times that are much shorter than the age of the Universe at this epoch.
    }
    \label{fig:tdepz11}
\end{figure}

\noindent{\textbf{Hypothesis i: Not Enough Early Halo Growth}}:
Several other studies have already shown that the rate of halo growth and the number density of halos massive enough to host the observed galaxies is sufficient in the standard $\Lambda$CDM cosmology with Planck-compatible parameters. \citet{Mason2022} showed that a model with ``maximally'' efficient star formation efficiency (all gas that accretes into halos forms stars with 100\% efficiency) can produce predicted UV LF that are consistent with or higher than the observational estimates up to $z \sim 14$.
Similarly, F23 showed that the \citet{Behroozi2015} model, which is based on a similar ansatz of star formation tracing halo accretion rates, can match the CEERS observed number densities out to $z=13$. In agreement with other studies, we therefore conclude that although it is still of course possible that dark matter and/or dark energy behave differently than assumed in the ``vanilla'' $\Lambda$CDM model, which would result in modified evolution of the formation rate of dark matter structures, this is not \emph{necessary} in order to accommodate the existing observational measurements up to $z\sim 14$.\\

\noindent{\textbf{Hypothesis ii: Inefficient Cooling}}:
The second hypothesis is that enough halos have collapsed and accreted baryons into their circumgalactic medium (CGM), but perhaps for some reason cooling from the CGM into the ISM is inefficient, so that not enough cold gas accumulates in the ISM to fuel adequate star formation. We examine this possibility in Fig.~\ref{fig:coolingfb}, which is shown for $z=11$. We see that in our models, the gas is cooling into the ISM at a rate similar to the accretion rate into the halo, so cooling is very efficient. The most common halos at these epochs have virial temperatures that place them near the peak of the cooling curve, leading to cooling times that are short compared to their dynamical times. Thus this is likely to be a fairly robust conclusion, in spite of our rather simplified treatment of cooling. \\

\noindent{\textbf{Hypothesis iii: Star formation is inefficient}}: Third, we consider the possibility that gas has cooled and accreted efficiently into the ISM, but perhaps for some reason galaxies are inefficient at making stars out of that gas in our models. For example, in our model we assume that star formation can only take place in gas that has been labelled as ``molecular'' by our gas phase partitioning scheme. We have adopted a partitioning scheme based on results presented by \citet{Gnedin2011}, in which molecular fraction depends on the gas surface density, the gas metallicity (which is assumed to trace the dust content), and the UV background (estimated from the local SFR). Thus at very low metallicity, for example, less \molh\ will be able to form, leading to a lower overall star formation efficiency \citep{Krumholz2009,Krumholz2011}. In addition, we assume that the star formation efficiency in \molh\ (or gas depletion time) is a function of the gas surface density, such that above a critical density of $\sim$ 70 $M_{\odot}$ pc$^{-2}$, the SFR density scales as the square of the \molh\ surface density. There is both theoretical and observational support for this assumed scaling \cite[see][and references therein]{Somerville2015}. In Fig.~\ref{fig:tdepz11} we show the depletion time for \molh\ only and for the total cold ISM gas (atomic, molecular, and ionized).  We see that, due to our adopted strong scaling of the SFE with gas surface density, and the very high predicted gas densities at this redshift, our models actually predict that the majority of galaxies at $z\sim 11$ (especially the more massive ones) have depletion times that are shorter than the age of the Universe at this epoch. We also found that decreasing the depletion time in our models by a factor of 2--10 did not significantly increase the predicted number of bright galaxies at $z\gtrsim 10$, confirming that the star formation efficiency is not the main rate limiting factor at this mass and redshift, at least in our modelling framework. \\

\noindent{\textbf{Hypothesis iv: Stellar Feedback is too strong}}:
Finally, we examine the impact of stellar feedback our predictions at $z \sim 11$. In our models, a mass outflow rate from the ISM is computed based on a parameterized mass loading factor. The mass loading factor is assumed to be only a function of the halo maximum circular velocity $V_{\rm max}$ and two free parameters, a normalization and a slope. These parameters are fixed to match $z=0$ observations (primarily the stellar mass function) and are not adjusted for the high redshift predictions. From Fig.~\ref{fig:coolingfb} (right panel), we can see that the predicted mass loading factors range from about 2 in the highest mass halos in our \gureft-90 box at $z=11$, to several hundred in the lower mass halos. This implies that for every solar mass per year of star formation, up to several hundred times that amount of gas is ejected from the galaxy and possibly from the halo. It seems clear that lower mass loadings for stellar driven winds would lead to more rapid build-up of stars and luminous galaxies at ultra-high redshifts. Thus we conclude that \emph{efficient stellar feedback} is the primary model ingredient that is responsible for the shortfall between the predicted and observed number density of luminous ($M_{\rm UV} \sim -19$ to $-20$) at $z\gtrsim 11$ in our models. Many other models and cosmological simulations similarly adopt efficient stellar and supernova driven winds to achieve the low observed stellar-to-halo mass ratios in low-mass halos in the nearby Universe.  \citet{Dekel2023} propose a promising explanation for why stellar feedback might be ineffective in the very early Universe, when the free-fall time is shorter than the time needed for massive stars to produce winds and supernovae.  We highlight this process as one that is crucial to investigate in more detail in future work.

\subsection{Theoretical and modelling uncertainties}

In the above sub-section, we attempted to quantify the roles of different physical processes \emph{within the framework of our existing models}. However, there are several other theoretical uncertainties that we have not touched on, which we summarize here. 

We note that the metallicity of the circumgalactic medium impacts how rapidly it cools, so chemical yields and the efficiency with which metals are ejected from, and circulate back into, the CGM will also impact the predicted cooling rates. Similarly, the chemical enrichment of the ISM and the formation of dust (not modelled here self-consistently) will have a strong impact on the modelling of star formation. We emphasize that the star formation prescription we have adopted was based on radiation-hydrodynamic simulations that are more than a decade old at this point, and on local universe conditions. In reality, \molh\ may simply trace the dense gas where star formation is efficient rather than being a pre-condition for star formation \citep[e.g.][]{Krumholz2011a}. 
At gas metallicities less than about 0.1 Z$_{\odot}$, additional drivers of the thermal state of the ISM become important, including reduced photoelectric heating, cosmic ray ionization heating, and \molh\ cooling \citep{Bialy2019}.

A related issue is the stochasticity, or burstyness, of star formation. If the process of star formation at early times is highly stochastic, then galaxies could experience short periods of time when their UV luminosities are boosted due to starbursts.  In fact, there is mounting observational evidence that ultra-high redshift galaxies did have extremely bursty star formation histories \citep[e.g.][]{Dressler2023, Endsley2023a, Looser2023}. Several modelling studies have shown that very bursty star formation could help to increase the number of luminous objects that would be observable at very high redshift and reduce the tension between model predictions and ultra-high redshift observations \citep{Mason2022,Shen2023a,Sun2023,Sun2023a}. We also find that adding a stochastic burst component in post-processing appears to be able to reconcile our fiducial model predictions with observations out to $z\sim 13$. However, in agreement with other studies \citep[e.g.][]{Shen2023a}, we find that a rather large stochastic component ($\sigma_{\rm UV} \sim 2$, where $\sigma_{\rm UV}$ is the root variance of a Gaussian random deviate in UV magnitude) is needed. This is much larger than the stochasticity produced in the high-resolution radiation-hydrodynamic cosmological simulations of \citet{Pallottini2023}, which yield typical $\sigma_{\rm UV} \sim 0.6$. We are therefore inclined to think that stochasticity alone is not responsible for resolving the discrepancy. However, weaker feedback, higher star formation efficiency, and bursty star formation are all expected to be consequences of the same physical changes in the ISM in ultra-high redshift galaxies. In future work, we plan to attempt to model these effects self-consistently in a physically motivated manner.

Another important process is the suppression of accretion and cooling by the meta-galactic photoionizing background after the Universe is reionized.  We adopt a highly simplified treatment of photoionization feedback in our models, in which we assume that the Universe was reionized everywhere instantaneously at $z=8$. After reionization, the baryon fraction that accretes into low mass halos is reduced, according to the the results of \citet{Okamoto2008}. However, in the real Universe, early forming, luminous objects in dense environments will create bubbles which will start to overlap and form larger reionized regions. We do not properly model the complex interplay between photo-ionization feedback from both local and meta-galactic radiation fields and stellar feedback. This will be increasingly important to model accurately as \textit{JWST} begins to push to lower luminosity galaxies. We also note here that \citet{Harikane2022} propose that the high observed number density of ultra-high redshift galaxies could be explained by the lack of suppression of star formation via photo-ionization before reionization. Our models do not include any suppression of star formation by the UV background at $z>8$, yet they still underpredict the observed counts --- therefore we suggest that this cannot be a complete explanation.

Another open question is how dust forms and is destroyed in the ISM at these early epochs. As noted above, dust is important for the chemistry and cooling of the ISM, and also impacts the observed SED of galaxies through attenuation and re-emission. Models and simulations have been presented that more self-consistently track the formation and destruction of dust through multiple channels \citep[e.g.][]{Popping2017,Vogelsberger2020a,Dayal2022}, but this physics is not incorporated into our current models. This will be an important ingredient to include as we update the realism of our modelling of star formation and the ISM in future work. 

Another possibility is that some of the UV light in some of the ultra-high-z galaxies could be contributed by an accreting supermassive black hole (AGN), which would provide a luminosity boost \citep{Pacucci2022}. Our current models include black hole seeding and accretion, and do not predict a significant contribution from AGN at these redshifts, but this is subject to the details of our seeding and accretion modelling. \citet{Volonteri2023} investigated this possibility, finding that a massive black hole seed (such as might form through the ``direct collapse'' mechanism) growing at close to the Eddington rate could be detectable with \textit{JWST}, and would be difficult to disentangle from star formation powered galaxies via the available \textit{JWST} colours. Candidate AGN at ultra-high redshift have been reported from early \textit{JWST} observations \citep[e.g.][]{Ono2022}. However, in current samples, the majority of the detected objects have extended (rather than point source) morphologies, suggesting that an AGN is not the dominant source of luminosity \citep[e.g.][]{Harikane2022}. 
Future \textit{JWST} observations will help to clarify the contribution of AGN to the detected ultra-high-z population, and place constraints on models of black hole seeding and accretion. 

Perhaps one of the largest uncertainties at ultra high redshifts is the properties of the stellar populations at these epochs. There is a large literature on the first stars (Pop III) which formed out of primordial gas and how their properties may have differed from present day stellar populations \citep[e.g.][]{Bromm1999,Abel2000,Zackrisson2011,Visbal2018}.
Although there is general agreement that the IMF was likely to have been more top-heavy in the early Universe, due both to the higher CMB temperature and lower metallicity, there is no detailed consensus on the expected Initial Mass Function (IMF) of Pop III stars, and even less on how the IMF may have evolved as the Universe is gradually polluted by metals (e.g. Pop III.2 stars). Clearly this will have a potentially profound impact on the galaxy SED, the chemical enrichment, and the stellar feedback --- none of the existing cosmological simulations that have been compared with the \textit{JWST} observations self-consistently incorporate these effects. \citet{Harikane2022} suggest that adopting a top-heavy IMF in early galaxies could help reconcile the theoretical predictions with observed number densities. They show that Pop III.1 and Pop III.2 populations with top-heavy IMFs modelled with the stellar population code {\sc YGGDRASIL} \citep{Zackrisson2011} can produce 3-4 times more UV photons for a given SFR than standard assumptions, in part due to the strong emission lines excited by massive stars.
\citet{Raiter2010} similarly show the expected change in the ratio of UV luminosity to SFR for several IMF variants that might arise in the conditions typical of the high redshift Universe, also finding an expected increase of a factor of 2-10.
This possibility has also been discussed by \citetalias{Finkelstein2022b}, who showed using abundance matching that a factor of $\sim 2$ boost in the UV luminosities could reconcile the differences with models. We showed a simple representation of how a top-heavy IMF would affect our model predictions in Fig.~\ref{fig:gureft_UVLF_var}, by simply boosting the luminosity of each galaxy in our models by a factor of 2--10. A boost of a factor of four brought our predicted number densities into good agreement with the observations. Clearly, modelling the formation of Pop III stellar populations and the transition through Pop III.2 to Pop II, and self-consistently incorporating appropriate stellar population models, is an important next step in modelling the ultra-high redshift Universe. Some early studies have found extremely blue UV slopes in ultra-high redshift galaxies, which could be an indication of the presence of extremely metal poor or even Pop III stars \citep{Topping2022}, although we note that these can also be indicative of high escape fraction \citep[e.g.][]{Endsley2023, Jung2023, Mascia2023, Mascia2023a}.
Emission line diagnostics from future \textit{JWST} spectroscopic observations will be able to shed light on the physical nature of the UV radiation sources \citep{Cleri2023}.

\section{Summary and Conclusions}
\label{sec:conclusions}

In this work, we presented predictions for galaxy populations at $z=8$--17 from the Santa Cruz semi-analytic model of galaxy formation, implemented within dark matter halo merger trees extracted from a new suite of N-body simulations called \gureft. 
\gureft\ is a suite of four cosmological volumes that can resolve halos over a broad range of mass ($5 \lesssim \log M_\text{h}/\text{M}_\odot \lesssim 12$). Furthermore, unlike many existing N-body simulations, \gureft\ stores finely spaced snapshots over the redshift range $40<z<6$ to ensure adequate temporal sampling of merger histories for galaxies out to ultra-high redshift. The resulting predictions have unprecedented dynamic range over the exciting $z\gtrsim 10$ redshift frontier that has recently begun to be probed by \textit{JWST}.

We presented predictions for the stellar mass function for $M_*$ ($3 \lesssim \log M_\text{*}/\text{M}_\odot \lesssim 10$) and for the \textit{intrinsic} (dust-free) rest-frame UV luminosity function for  $-9 \gtrsim \log M_\text{UV} \gtrsim -24$ between $z = 8$ to 17. We also presented predictions for the scaling relations between stellar mass $M_*$, rest-frame UV luminosity $M_\text{UV}$, UV slope $\beta_\text{UV}$, and gas phase metallicity $Z_{\rm gas}$ at $z=8$, 11, and 13. In an appendix, we presented predictions for the relationship between stellar mass and rest-UV luminosity and halo mass at these same redshifts. These model outputs are compared to recent \textit{JWST} and selected past \textit{Hubble} and \textit{Spitzer} observations. We also presented a detailed discussion of the observational uncertainties, including the potential impact of cosmic variance. We reach the following main conclusions:
\begin{itemize}
    \item We find excellent agreement between our predicted UV LF and observational estimates at $z\sim 8$--10. At $z\sim 11$, our predicted UVLF is about a factor of 10-15 lower than observational estimates at $M_{\rm UV}\simeq -20$. At $z=13$, our model predictions at $M_{\rm UV}\simeq -20$ are about a factor of 30 lower than the observational estimates. 
    
    \item We estimate representative uncertainties in the observed number counts at $z=11$ and $z=14$ due to cosmic variance from large scale structure. We find that this can be a significant source of uncertainty, and that the estimates themselves are highly uncertain, due to our lack of knowledge about the host halo masses of the observed galaxies and their clustering.

    \item We find that a modest ``boost'' in the ratio between UV luminosity and stellar mass of about a factor of four would bring our model predictions into reasonable agreement with the existing observational constraints at $z=11$ and $z=13$. This is consistent with what might be expected from a top-heavy IMF. 

    \item We find that adding a stochastic ``burst'' component to star formation in post-processing can also bring our fiducial models into agreement with the observations, but a potentially unphysically large stochastic component is required.
    
    \item Our models predict fairly strong evolution in the stellar mass function and the UVLF over the redshift interval $8$--12. We predict a factor of 250 evolution at a stellar mass of $10^9 M_{\odot}$, and a factor of $\sim 200$ at $M_{\rm UV}=-20$, between redshift 12 and 8. Moreover, the redshift evolution for luminous (massive) galaxies is more rapid than that for less luminous (massive) galaxies.
    
    \item We find very good agreement between our model predictions for the stellar mass vs. $M_{\rm UV}$ relation and those derived from SED fitting to observations at $z=8$--13. 
    
    \item We find that our predicted rest-UV slopes lie close to the medians of the UV slopes derived from SED fitting to observations at $z=8$--13, but that the observed slopes have much more diversity than our model predictions, with the observed UV slopes extending both to redder and much bluer values. This could indicate the presence of dust and of Pop III type stellar populations in the observed samples.  
    
    \item The ISM metallicities in our models range from 0.01-0.1 $Z_{\odot}$ at $z=13$, and reach solar values in the most massive galaxies in our simulated volume at $z=8$. Our models predict an interesting ``inverted'' evolution in $Z_{\rm gas}$, with $Z_{\rm gas}$ at a fixed stellar mass \emph{decreasing} with cosmic time over the time interval $z=17$ to $z=8$.

    \item Our models predict very weak evolution in the scaling between stellar mass and halo mass, and between $M_{\rm UV}$ and halo mass, over the redshift interval $z=13$--8. If correct, this implies that the observed evolution of the SMF and UV LF is closely tracing the evolution of the underlying dark matter halo mass function. 

    \item We investigate which of the physical processes in our models is most likely to be responsible for the much lower number density of predicted galaxies at $z=11$--13 compared with observational estimates. We consider the growth rate of halos predicted by vanilla $\Lambda$CDM, and the efficiency of cooling, the conversion of dense gas into stars, and the stellar feedback. We conclude that the very strong stellar feedback adopted in our models is the most likely ingredient that could be modified to yield higher predicted galaxy number densities at ultra-high redshifts.
    
\end{itemize}

Ongoing observations with \textit{JWST} and other facilities such as ALMA will soon yield more robust constraints on the redshifts and number densities of ultra-high redshift galaxies, as well as additional information about these objects such as their metallicities and the conditions in the ISM. These observations will provide an unprecedented laboratory for developing a new generation of theoretical models of galaxy formation, which will incorporate more physically grounded and self-consistent treatments of key processes such as star formation, stellar feedback, stellar populations, and black hole seeding, accretion and feedback.

%%%%%%%%%%%%%%%%%%%%%%%%%%%%%%%%%%%%%%%%%%%%%%%%%%
\section*{Data Availability}

Catalogs of the galaxy properties for the predictions presented in this paper are available in the Data Product Portal hosted by the Flatiron Institute at \url{http://flathub.flatironinstitute.org}.

\section*{Acknowledgements}

We thank Eli Visbal, David Spergel, and the CCA galaxy formation group for useful discussions. We thank Christian Jespersen for suggestions and comments that improved the manuscript.
The \gureft\ simulation suite and Santa Cruz Semi-Analytic galaxy formation model was run on the Flatiron Institute computing cluster \textit{rusty}, managed by the Scientific Computing Core (SCC).
AY is supported by an appointment to the NASA Postdoctoral Program (NPP) at NASA Goddard Space Flight Center, administered by Oak Ridge Associated Universities under contract with NASA.
RSS acknowledges support from the Simons Foundation. 
Support from program numbers ERS-01345 and AR-02108 was provided through a grant from the Space Telescope Science Institute (STScI) under NASA contract NAS5-03127.

%%%%%%%%%%%%%%%%%%%%%%%%%%%%%%%%%%%%%%%%%%%%%%%%%%



\bibliographystyle{mnras}
\bibliography{ultra-z-galaxies} % if your bibtex file is called example.bib

%\begin{thebibliography}{99}
%\end{thebibliography}

%%%%%%%%%%%%%%%%%%%%%%%%%%%%%%%%%%%%%%%%%%%%%%%%%%

%%%%%%%%%%%%%%%%% APPENDICES %%%%%%%%%%%%%%%%%%%%%

\appendix

\section{Tabulated values for HMFs, SMFs and UV LFs }
\label{sec:tabulated}

Tabulated ultra-$z$ HMFs, SMFs, and UV LFs are provided in Tables \ref{tab:HMF_tabulated}, \ref{tab:SMF_tabulated}, and \ref{tab:UVLF_tabulated}. These distribution functions are constructed with halos and galaxies in mass range well-resolved in the four \gureft\ simulated volumes.

\begin{table*}
	\centering
	\caption{Tabulated HMFs at z = 8 -- 20}
	\label{tab:HMF_tabulated}
	\begin{tabular}{lccccccccccccccccc} % four columns, alignment for each
		\hline
                      & \multicolumn{13}{c}{$\log_\text{10}(\phi[\text{dex}^{-1} \text{Mpc}^{-3}]) $}\\
		$\log (M_\text{h}/\text{M}_\odot)$ & $z = 8$ & $z = 9$ & $z = 10$ & $z = 11$ & $z = 12$ & $z = 13$ & $z = 14$ & $z = 15$ & $z = 16$ & $z = 17$ & $z = 18$ & $z = 19$ & $z = 20$\\
		\hline
5.0 & 3.26 & 3.23 & 3.19 & 3.14 & 3.08 & 3.0 & 2.92 & 2.83 & 2.75 & 2.63 & 2.52 & 2.39 & 2.28\\
5.5 & 3.58 & 3.54 & 3.48 & 3.42 & 3.34 & 3.25 & 3.15 & 3.04 & 2.94 & 2.8 & 2.67 & 2.52 & 2.38\\
6.0 & 3.19 & 3.14 & 3.08 & 3.0 & 2.91 & 2.81 & 2.69 & 2.57 & 2.45 & 2.29 & 2.14 & 1.96 & 1.8\\
6.5 & 2.7 & 2.63 & 2.56 & 2.48 & 2.36 & 2.25 & 2.11 & 1.96 & 1.83 & 1.64 & 1.46 & 1.24 & 1.06\\
7.0 & 2.18 & 2.11 & 2.02 & 1.91 & 1.77 & 1.63 & 1.46 & 1.28 & 1.12 & 0.9 & 0.65 & 0.37 & 0.13\\
7.5 & 1.65 & 1.55 & 1.44 & 1.3 & 1.14 & 0.95 & 0.74 & 0.59 & 0.34 & -0.01 & -0.23 & -0.5 & -0.64\\
8.0 & 1.09 & 0.97 & 0.83 & 0.68 & 0.47 & 0.26 & 0.02 & -0.24 & -0.48 & -0.8 & -1.13 & -1.58 & -1.88\\
8.5 & 0.53 & 0.37 & 0.18 & -0.02 & -0.26 & -0.53 & -0.84 & -1.2 & -1.48 & -1.79 & -2.23 & -2.53 & -2.73\\
9.0 & -0.1 & -0.3 & -0.53 & -0.79 & -1.1 & -1.43 & -1.82 & -2.23 & -2.62 & -3.05 & -3.31 & -3.84 & -4.24\\
9.5 & -0.76 & -1.03 & -1.38 & -1.71 & -2.07 & -2.45 & -2.9 & -3.36 & -3.63 & -4.24 & nan & nan & nan\\
10.0 & -1.53 & -1.88 & -2.28 & -2.68 & -3.21 & -3.79 & -4.37 & -5.16 & -5.29 & nan & nan & nan & nan\\
10.5 & -2.4 & -2.84 & -3.39 & -3.87 & -4.42 & -5.47 & -5.77 & nan & nan & nan & nan & nan & nan\\
11.0 & -3.46 & -3.95 & -4.65 & -5.29 & nan & nan & nan & nan & nan & nan & nan & nan & nan\\
11.5 & -4.62 & -5.77 & nan & nan & nan & nan & nan & nan & nan & nan & nan & nan & nan\\
		\hline
	\end{tabular}
\end{table*}

\begin{table*}
	\centering
	\caption{Tabulated SMFs at z = 8 -- 17}
	\label{tab:SMF_tabulated}
	\begin{tabular}{lcccccccccccccc} % four columns, alignment for each
		\hline
                      & \multicolumn{10}{c}{$\log_\text{10}(\phi[\text{dex}^{-1} \text{Mpc}^{-3}]) $}\\
		$\log (M_*/\text{M}_\odot)$ & $z = 8$ & $z = 9$ & $z = 10$ & $z = 11$ & $z = 12$ & $z = 13$ & $z = 14$ & $z = 15$ & $z = 16$ & $z = 17$\\
		\hline
3.0 & 0.06 & 0.04 & 0.01 & -0.04 & -0.16 & -0.29 & -0.45 & -0.55 & -0.75 & -1.12\\
3.5 & 0.27 & 0.23 & 0.14 & 0.0 & -0.17 & -0.37 & -0.6 & -0.88 & -1.08 & -1.4\\
4.0 & 0.26 & 0.14 & -0.02 & -0.21 & -0.46 & -0.67 & -0.98 & -1.25 & -1.51 & -1.95\\
4.5 & 0.01 & -0.16 & -0.32 & -0.52 & -0.76 & -1.02 & -1.38 & -1.68 & -1.97 & -2.27\\
5.0 & -0.29 & -0.46 & -0.66 & -0.88 & -1.16 & -1.43 & -1.75 & -2.07 & -2.36 & -2.85\\
5.5 & -0.62 & -0.8 & -1.03 & -1.28 & -1.59 & -1.88 & -2.27 & -2.69 & -3.05 & -3.33\\
6.0 & -0.96 & -1.19 & -1.44 & -1.71 & -2.06 & -2.41 & -2.88 & -3.24 & -3.52 & -4.06\\
6.5 & -1.33 & -1.57 & -1.85 & -2.18 & -2.64 & -3.0 & -3.36 & -3.93 & -4.06 & -4.84\\
7.0 & -1.69 & -2.0 & -2.32 & -2.73 & -3.14 & -3.5 & -3.93 & -4.36 & nan & nan\\
7.5 & -2.12 & -2.45 & -2.85 & -3.27 & -3.58 & -3.99 & nan & nan & nan & nan\\
8.0 & -2.57 & -2.99 & -3.54 & -3.66 & -4.54 & nan & nan & nan & nan & nan\\
8.5 & -3.08 & -3.57 & -4.13 & -4.58 & -4.84 & nan & nan & nan & nan & nan\\
9.0 & -3.67 & -4.2 & -4.75 & -5.37 & -6.07 & nan & nan & nan & nan & nan\\
9.5 & -4.3 & -4.95 & -6.07 & nan & nan & nan & nan & nan & nan & nan\\
10.0 & -4.99 & -5.59 & nan & nan & nan & nan & nan & nan & nan & nan\\
10.5 & -6.07 & nan & nan & nan & nan & nan & nan & nan & nan & nan\\
		\hline
	\end{tabular}
\end{table*}

\begin{table*}
	\centering
	\caption{Tabulated UV LFs at z = 8 -- 17}
	\label{tab:UVLF_tabulated}
	\begin{tabular}{lcccccccccccccc} % four columns, alignment for each
		\hline
                      & \multicolumn{10}{c}{$\log_\text{10}(\phi[\text{mag}^{-1} \text{Mpc}^{-3}]) $}\\
		$M_\text{UV}$ & $z = 8$ & $z = 9$ & $z = 10$ & $z = 11$ & $z = 12$ & $z = 13$ & $z = 14$ & $z = 15$ & $z = 16$ & $z = 17$\\
		\hline
-12.0 & -1.09 & -1.27 & -1.44 & -1.56 & -1.84 & -2.09 & -2.46 & -2.71 & -2.99 & -3.61\\
-12.5 & -1.25 & -1.39 & -1.55 & -1.68 & -2.02 & -2.32 & -2.65 & -3.0 & -3.31 & -3.72\\
-13.0 & -1.33 & -1.52 & -1.74 & -1.92 & -2.21 & -2.44 & -2.91 & -3.24 & -3.58 & -3.66\\
-13.5 & -1.5 & -1.69 & -1.92 & -2.05 & -2.42 & -2.64 & -3.29 & -3.5 & -3.69 & -4.24\\
-14.0 & -1.61 & -1.83 & -2.09 & -2.24 & -2.6 & -2.89 & -3.36 & -3.69 & -3.88 & -4.84\\
-14.5 & -1.76 & -2.0 & -2.23 & -2.46 & -2.82 & -3.09 & -3.5 & -3.76 & -4.24 & -4.84\\
-15.0 & -1.91 & -2.14 & -2.46 & -2.67 & -3.11 & -3.28 & -3.76 & -4.54 & -4.36 & -4.84\\
-15.5 & -2.08 & -2.3 & -2.59 & -2.93 & -3.28 & -3.58 & -4.06 & -4.36 & -4.54 & nan\\
-16.0 & -2.24 & -2.46 & -2.84 & -3.11 & -3.31 & -3.63 & -4.36 & -4.36 & nan & nan\\
-16.5 & -2.43 & -2.61 & -3.02 & -3.35 & -3.76 & -4.06 & -4.24 & nan & -5.77 & nan\\
-17.0 & -2.49 & -2.86 & -3.24 & -3.54 & -3.88 & -4.24 & -5.07 & -5.77 & -6.07 & -5.77\\
-17.5 & -2.73 & -3.19 & -3.44 & -3.69 & -4.14 & -4.24 & -5.77 & -5.77 & -6.07 & nan\\
-18.0 & -2.85 & -3.3 & -3.73 & -4.03 & -4.59 & -4.84 & -5.59 & -5.77 & -6.07 & nan\\
-18.5 & -3.11 & -3.55 & -4.03 & -4.52 & -4.75 & -5.29 & -5.77 & -6.07 & nan & nan\\
-19.0 & -3.32 & -3.78 & -4.35 & -4.67 & -5.47 & -5.59 & nan & nan & nan & nan\\
-19.5 & -3.58 & -4.09 & -4.51 & -4.84 & -5.37 & nan & -6.07 & nan & nan & nan\\
-20.0 & -3.76 & -4.34 & -4.79 & -5.16 & -6.07 & -6.07 & nan & nan & nan & nan\\
-20.5 & -4.16 & -4.54 & -5.11 & -6.07 & -6.07 & nan & nan & nan & nan & nan\\
-21.0 & -4.32 & -4.81 & -5.47 & -6.07 & nan & nan & nan & nan & nan & nan\\
-21.5 & -4.58 & -5.22 & -6.07 & -6.07 & nan & nan & nan & nan & nan & nan\\
-22.0 & -4.99 & -5.59 & nan & nan & nan & nan & nan & nan & nan & nan\\
-22.5 & -5.16 & -5.47 & nan & nan & nan & nan & nan & nan & nan & nan\\
-23.0 & -5.37 & nan & nan & nan & nan & nan & nan & nan & nan & nan\\
		\hline
	\end{tabular}
\end{table*}

\section{Dark matter halo connection}
\label{sec:halo_connection}

\begin{figure}
    \includegraphics[width=\columnwidth]{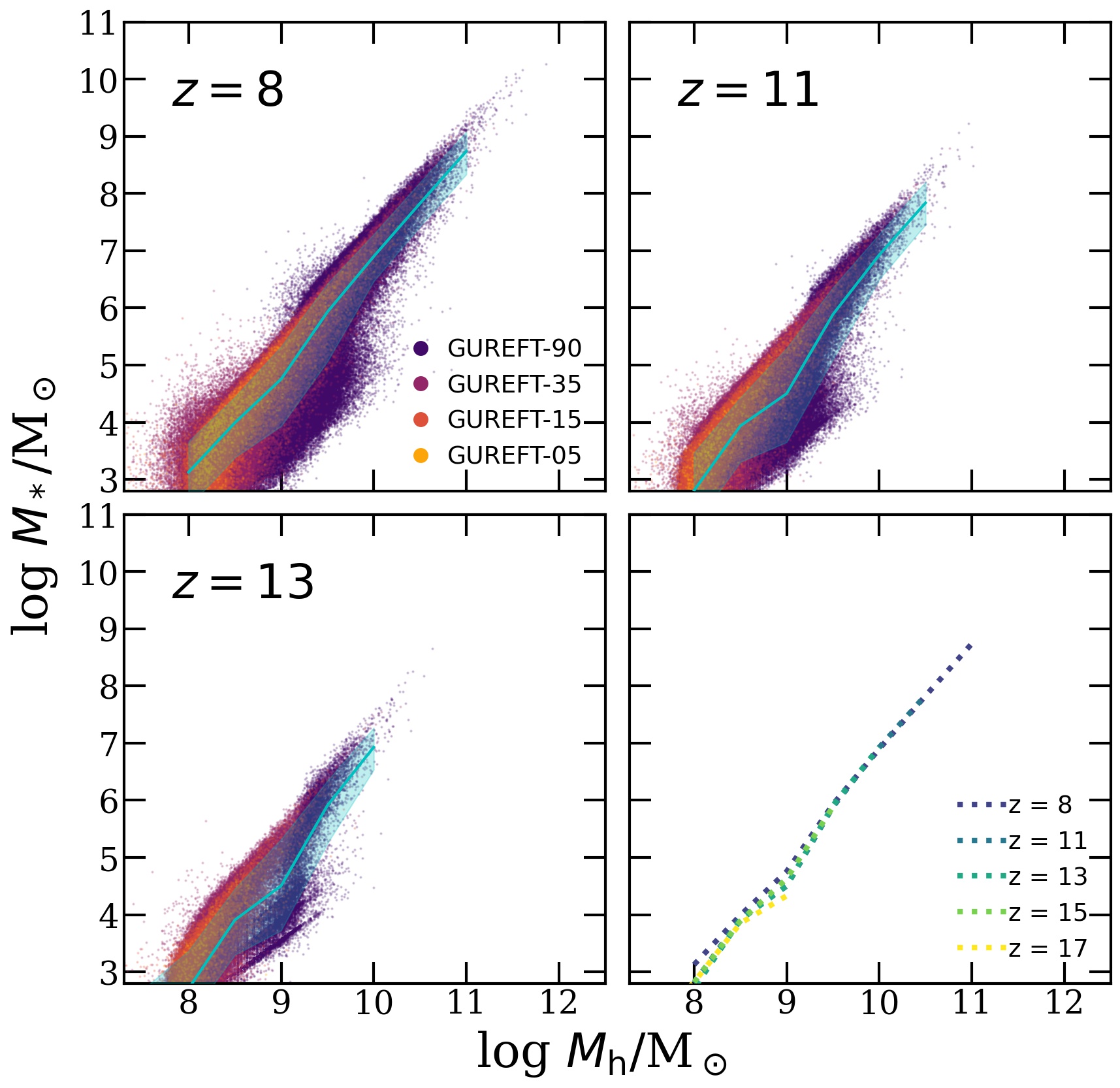}
    \caption{
        Stellar mass vs. halo mass ($M_*$--$M_\text{h}$) relation at $z = 8$, 11, and 13. The data points show SAM galaxies from the \gureft\ boxes. 
        The cyan line marks the median and shaded area marks the 16th and 84th percentiles for galaxies in bins of $M_\text{h}$. The last panel summarizes the evolution of $M_*$--$M_\text{h}$ relation.
    }
    \label{fig:4panel_mh_ms}
\end{figure}

\begin{figure}
    \includegraphics[width=\columnwidth]{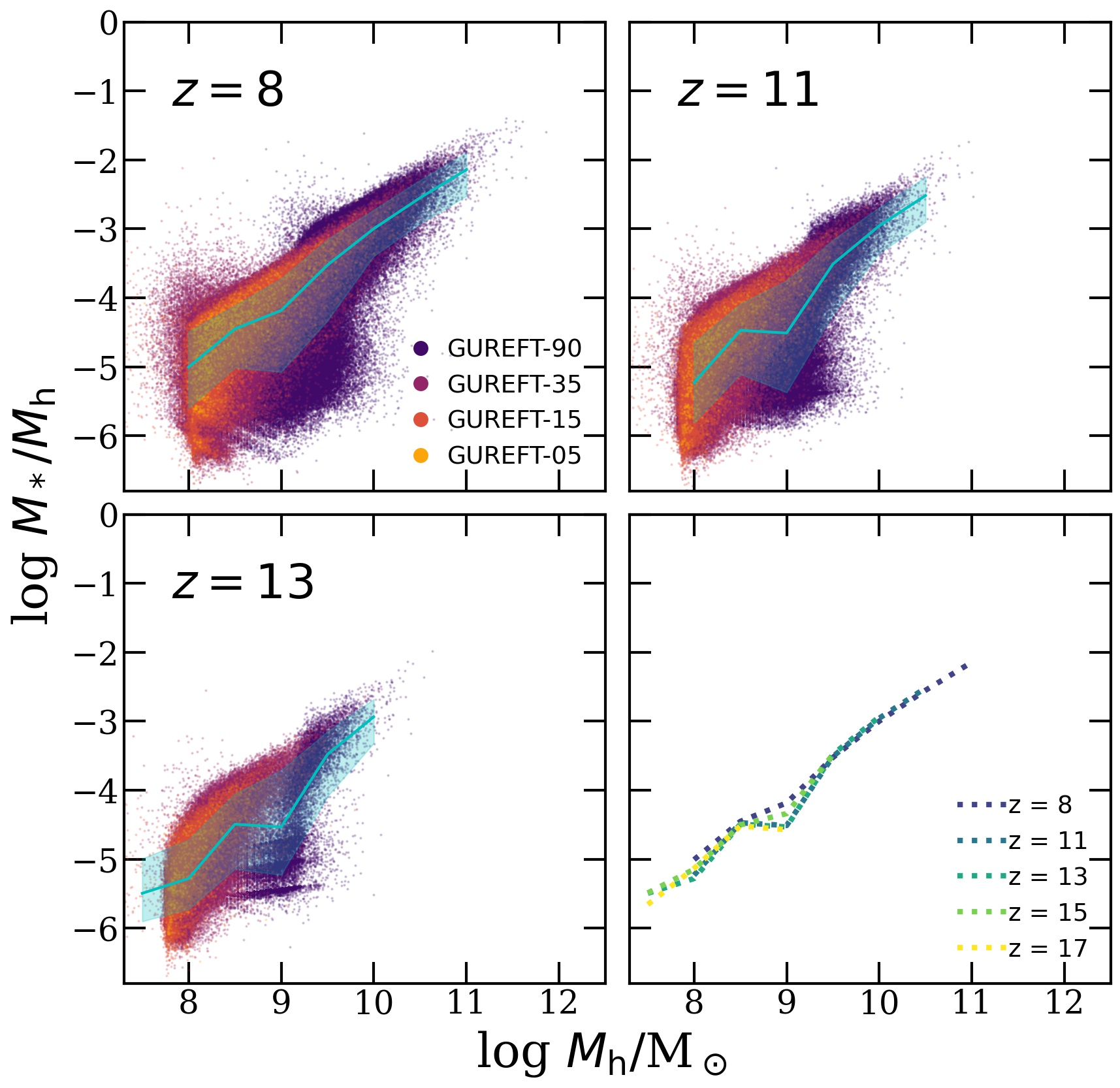}
    \caption{
        The stellar-to-halo mass ratio (SHMR) at $z = 8$, 11, and 13. The data points show SAM galaxies from the \gureft\ boxes. 
        The cyan line marks the median and shaded area marks the 16th and 84th percentiles for galaxies in bins of $M_\text{h}$. The last panel summarizes the evolution of SHMR.
    }
    \label{fig:4panel_mh_shmr}
\end{figure}

\begin{figure}
    \includegraphics[width=\columnwidth]{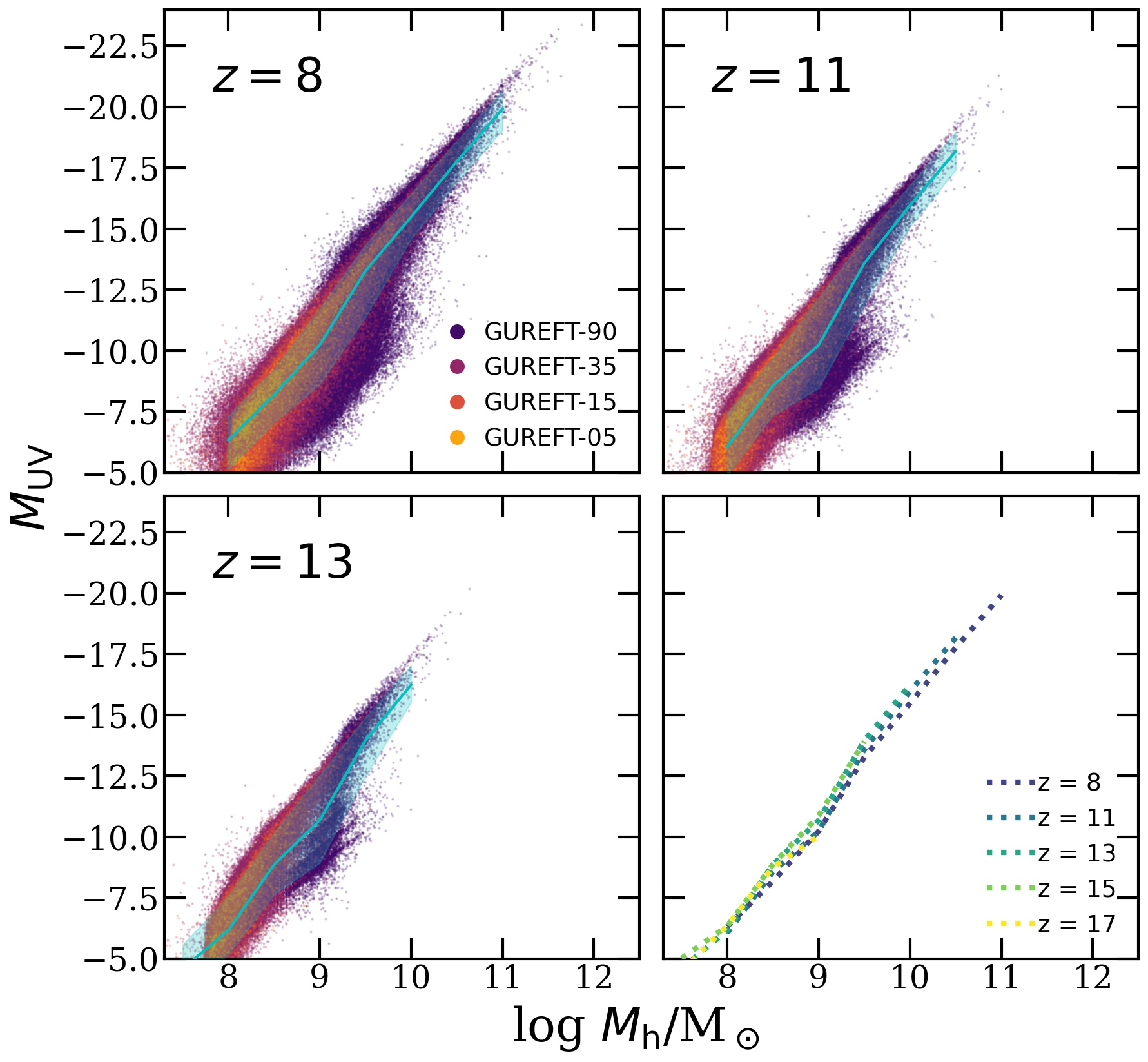}
    \caption{
        Rest UV magnitude vs. halo mass ($M_\text{UV}$--$M_\text{h}$) relation at $z = 8$, 11, and 13. The data points show SAM galaxies from the \gureft\ boxes. 
        The cyan line marks the median and shaded area marks the 16th and 84th percentiles for galaxies in bins of $M_\text{h}$. The last panel summarizes the evolution of $M_\text{UV}$--$M_\text{h}$ relation.
    }
    \label{fig:4panel_mh_muv}
\end{figure}

We show several commonly used scaling relations that connect galaxy properties with halo mass. Figs.~\ref{fig:4panel_mh_ms}, \ref{fig:4panel_mh_shmr}, and \ref{fig:4panel_mh_muv} show stellar mass, stellar-to-halo mass ratio, and rest-frame $M_\text{UV}$ versus halo mass.

%%%%%%%%%%%%%%%%%%%%%%%%%%%%%%%%%%%%%%%%%%%%%%%%%%

% Don't change these lines
\bsp	% typesetting comment
\label{lastpage}
\end{document}